\documentclass[12pt,twoside]{report}
\usepackage{amsmath,amssymb,svcon2e}
\usepackage{hyperref}  
\usepackage[numbers,sort&compress]{natbib}
\usepackage{hypernat} 
\catcode`@=12
\makeatletter
\def\@crosshairs{\vbox to 0pt{}}

\renewcommand{\theequation}{\arabic{section}.\arabic{equation}}

\newcommand{\address}[1]{{\em #1\\ }}

\renewcommand{\cases}[1]{\left\{\begin{array}{ll} #1 \end{array}\right.}
\newcommand{\comment}[1]{}

\renewcommand{\arraystretch}{0.91}

\def\swnedots{\mathinner{\mkern1mu\raise1pt\vbox{\kern7pt\hbox{.}}\mkern2mu
\raise4pt\hbox{.}\mkern2mu\raise7pt\hbox{.}\mkern1mu}}
\def\({\left(}
\def\){\right)}
\def\<{\langle}
\def\>{\rangle}

\newcommand{\ds}{\displaystyle}

\newcommand{\mi}{\!-\!}
\newcommand{\plus}{\!+\!}

\def\topped#1#2{\genfrac{}{}{0pt}{3}{#1}{#2}}

\font\tenmsb=msbm10 scaled \magstep 2
\font\sevenmsb=msbm10 scaled \magstep 1
\font\fivemsb=msbm10
\newfam\msbfam
\textfont\msbfam=\sevenmsb
\scriptfont\msbfam=\sevenmsb
\scriptscriptfont\msbfam=\fivemsb

\newfam\smmsbfam
\textfont\smmsbfam=\sevenmsb

\newfam\Bmsbfam
\textfont\Bmsbfam=\tenmsb

\font\bigtenmsb=msbm10 scaled \magstep 3
\font\bigsevenmsb=msbm7 scaled \magstep 3
\font\bigfivemsb=msbm5 scaled \magstep 3
\newfam\bigmsbfam
\textfont\bigmsbfam=\bigtenmsb
\scriptfont\bigmsbfam=\bigsevenmsb
\scriptscriptfont\bigmsbfam=\bigfivemsb

\newcommand{\W}[2]{\!\left(\ds\left.\!\begin{array}{cccccc}#1%
\end{array}\right #2\right)}

\newcommand{\B}[7]{B^{#1}\!\!\left(\left.\!#4
\begin{array}{@{\;\,}c@{\;\,}c@{\:}}#5&#6\\
#2&#3\end{array}
\right|#7\right)}
\renewcommand{\ss}{\scriptstyle}
\newcommand{\sss}{\scriptscriptstyle}
\newcommand{\half}{{\sss1\!/\!2}}

\newcommand{\p}[2]{\makebox(0,0)[#1]{$#2$}}
\newcommand{\pp}[2]{\makebox(0,0)[#1]{$\ss#2$}}

\newcommand{\textB}[6]{\begin{picture}(#1,#2)\put(#3,#4){\p{#5}{%
\ds#6} 
}\end{picture}}
\renewcommand{\vec}[1]{\mbox{\boldmath$#1$}}

\renewcommand{\B}[2]{\!\!\left(\left.\!\!\!
\renewcommand{\arraystretch}{.3}
	\begin{array}{c@{\,}c@{\,}c@{\,}c}
     #1
	\end{array}\!
	\right#2\right)}

\begin{document}
\pagenumbering{arabic}
    \thispagestyle{empty}
\chapter{Integrable Boundaries and Universal TBA
Functional Equations}
\author{C. H. Otto Chui, Christian Mercat
and Paul A. Pearce}
\chapterauthors{C. H. Otto Chui, Christian Mercat
and Paul~A. Pearce\footnote{
\href{mailto:C.Chui@ms.unimelb.edu.au}{C.Chui},
\href{mailto:C.Mercat@ms.unimelb.edu.au}{C.Mercat},
\href{mailto:P.Pearce@ms.unimelb.edu.au}{P.Pearce@ms.unimelb.edu.au}\quad
This article is dedicated to Prof. Barry McCoy on the occasion of his sixtieth
birthday.}}
\noindent\address{Department of Mathematics and Statistics\\
University of Melbourne\\Parkville, Victoria 3010, Australia}
\begin{abstract}
We derive the fusion hierarchy of functional equations for critical
$A$-$D$-$E$ lattice models related to the $s\ell(2)$ unitary minimal
models, the parafermionic models and the supersymmetric models of
conformal field theory and deduce the related TBA functional
equations.  The derivation uses fusion projectors and applies in the
presence of all known integrable boundary conditions on the torus and
cylinder.  The resulting TBA functional equations are {\em universal }
in the sense that they depend only on the Coxeter number of the
$A$-$D$-$E$ graph and are independent of the particular integrable
boundary conditions.  We conjecture generally that TBA functional
equations are universal for all integrable lattice models associated
with rational CFTs and their integrable perturbations.
\end{abstract}
\section{Introduction}
\label{sec:Introduction}
Like all good scientists, Barry McCoy has long since appreciated the
power and the beauty of universality in physics and its implications
in mathematics.  This is evident starting with his work on the Ising
model~\cite{McCoyWu} and continues through to his introduction of
Universal Chiral Partition Functions~\cite{UCPF1,UCPF2}.  In this
article we follow McCoy's lead and study the universality of TBA
functional equations.

Ever since Baxter solved~\cite{Baxt8V} the eight-vertex model,
commuting transfer matrix functional
equations~\cite{BaxP82,Pea87,BR89,Pea92,KP92,KNS1} 
have been at the heart of the exact solution of two-dimensional
lattice models on a periodic lattice by Yang-Baxter
methods~\cite{Bax}.  For theories such as the $A$-$D$-$E$ models
considered here, these equations provide the key to obtaining free
energies, correlation lengths and finite-size corrections.  At
criticality, the finite-size corrections are related to the central
charges and scaling dimensions of the associated conformal field
theory (CFT).  Off-criticality, these corrections yield the scaling
energies of the associated (perturbed) integrable quantum field theory
(QFT).  The fundamental form of the functional equations involves
fusion of the Boltzmann weights on the lattice and reflect the fusion
rules of the associated CFT. However, in order to solve for
finite-size corrections these functional equations need to be recast
in the form of a $Y$-system or
TBA functional equations~\cite{Zam1,
KP92,PN98}.
Miraculously, it is then possible to solve~\cite{KP92} for the central
charges and scaling dimensions using some special tricks and
dilogarithm identities~\cite{Kir93}.

More recently, it has been realized~\cite{Skl88,BPO'B96} that the
Yang-Baxter methods and functional equations can be extended to
systems in the presence of integrable boundaries on the cylinder by
working with double row transfer matrices.  It is then possible to
calculate surface free energies and interfacial tensions~\cite{O'BP97}
as well as finite-size corrections and conformal partition
functions~\cite{O'BPW97}.  The critical $A$-$D$-$E$ models correspond,
for different choices of regimes and/or fusion level, to unitary
minimal models~\cite{BP01}, parafermion theories~\cite{MP} and
superconformal theories~\cite{RP01}.  These theories include the
critical Ising, tricritical Ising and critical 3-state Potts models. 
For these theories, an integrable boundary condition on the cylinder
can be constructed for each allowed conformal boundary
condition~\cite{BPPZ00}.  It is also possible to
construct~\cite{CMOP1} integrable seams for each conformal twisted
boundary condition~\cite{PZ0011021} on the torus.  In all such cases
it should be possible to obtain the {\em universal} conformal
properties in the continuum scaling limit by solving suitable
functional equations.

In this paper we derive general fusion and TBA functional equations
for the critical \mbox{$A$-$D$-$E$} lattice models.  Although the
fusion hierarchy of functional equations is not universal, we show in
this paper that the $Y$-system or TBA functional equations for the
$A$-$D$-$E$ models are \emph{universal} in the sense that they depend
on the $A$-$D$-$E$ graph only through its Coxeter number, and more
importantly, they are independent of the choice of integrable boundary
conditions.

The universality of the TBA equations has important consequences.  It
asserts that the functional equations are the same for all twisted
boundaries on the torus and open boundaries on the cylinder. 
Therefore the same functional equations must be solved in all cases! 
So no new miracles, beyond the periodic case, are required to solve
these equations in the presence of conformal boundaries.  Instead, the
different solutions required among the infinite number of possible
solutions to the TBA functional equations are selected by appropriate
analyticity requirements.  These analyticity properties allow for the
derivation of non-linear integral equations (NLIE) that can be solved
for the complete spectra of the transfer matrices and the universal
conformal data encoded in the finite-size corrections.  Of course the
analyticity properties are not universal.  However, one strength of
the lattice approach is that the analyticity determined by the
structure of zeros and poles of the eigenvalues of the transfer
matrices can be probed directly by numerical calculations on
finite-size lattices.  In this way it is possible to build up case by
case a complete picture of the required analyticity properties.

The layout of the paper is as follows.  We first recall some results
about fused $A$-$D$-$E$ models in
Sections~\ref{sec:FaceWeights}--\ref{sec:FusedFace}.  In
Section~\ref{sec:Transfer}, we define the transfer matrix for the
different boundary conditions, on the torus and on the cylinder, with
and without seams.  In Section~\ref{sec:FusionHierarchies}, we state
the main result of the paper, that is the TBA equation, the boundary
specific functional equations and their universal form.  In
Section~\ref{sec:Derivation}, we derive the TBA and related functional
equations.  We first study the general idea which is based on local
properties in~\ref{sec:LocalProperties} and we then apply it to the
torus in~\ref{sec:FunctEqTorus} and the cylinder
in~\ref{sec:FunctEqCylinder}.  We conclude with a discussion in
Section~\ref{sec:Discussion}.

The methods developed in this paper should extend to the general
$s\ell(2)$ coset models.  However, we focus our attention on the
unitary minimal, superconformal and parafermionic series,
corresponding to fusion levels $p=1, 2$ and negative regime
respectively, because only in these cases is our knowledge of the
integrable and conformal boundary conditions complete.

Our results can also be extended, using the methods of \cite{BPO'B96},
to the $A$ and $D$ lattice models off-criticality yielding precisely
the same TBA functional equations.  In these cases the lattice models
admit elliptic solutions to the Yang-Baxter equation where the
elliptic nome plays the role of the deviation from critical
temperature.  Although the TBA functional equations are traditionally
associated with the thermodynamics of quantum spin chains at finite
temperature $T$ they are derived here from a two-dimensional lattice
approach.  It is of course well known that the relevant quantum spin
chains can be obtained as a logarithmic derivative of the transfer
matrices with respect to the spectral parameter.  In fact, the two
approaches based on spin chains and two-dimensional lattice models are
entirely equivalent~\cite{BR89,Klump}.  Most importantly, the TBA
equations originally conjectured by
Zamolodchikov~\cite{Zam1} 
can be derived~\cite{PN98} within the lattice approach yielding the
precise relation between the temperature $T$ and the elliptic nome of
the lattice model.  The critical case that we focus on here just
corresponds to $T=0$ and vanishing elliptic nome.

It is worthwhile to mention possible applications of the TBA
functional equations.  Given prescribed conformal boundary conditions,
the TBA functional equations can of course be used to derive
non-linear integral equations (NLIE) which in principle can be solved
for the known conformal partition functions.  Of more interest,
however, is to include in the NLIE the effect of a boundary field
$\xi$ to perturb away from the conformal boundary.  This induces an
integrable boundary flow and allows for the study of integrable
boundary flows between distinct conformal boundary
conditions~\cite{FeveratiPR}.  In the off-critical case for the $A$
and $D$ models, a similar analysis~\cite{PCA} allows for the study in
the presence of boundaries of thermal renormalization group flows
connecting different coset models.

\subsection{Face Weights}
\label{sec:FaceWeights}
A lattice model in the $A$-$D$-$E$ series is associated with a graph
$G$, of $A$, $D$ or $E$ type.  The spins are nodes of the graph $G$ and
neighbouring sites on the lattice must be neighbouring nodes of
the graph.  The probability distribution of spins is defined by the
critical (unfused) Boltzmann weight of each face (or plaquette) of
spins, depending on a spectral parameter $u$:
\begin{equation}
     W^{11}\W{d&c\\a&b}{|u}=
     \setlength{\unitlength}{.5cm}
\begin{picture}(3.5,1)(-.6,-.2)
\put(0,-1){\framebox(2,2){\p{}{u}}}
\put(.1,-.9){\pp{bl}{\searrow}}
\put(-.25,-1.25){\pp{}{a}}
\put(2.25,-1.25){\pp{}{b}}
\put(2.25,1.25){\pp{}{c}}
\put(-.25,1.25){\pp{}{d}}
\end{picture}
=s(\lambda-u)\delta_{ac}+
     s(u)\sqrt{\frac{\psi_{a}\psi_{c}}{\psi_{b}\psi_{d}}}\delta_{bd}
            \label{eq:W11}
\end{equation}
where $g$ is the Coxeter number of $G$, $\lambda=\frac{\pi}{g}$,
$s(u)=\frac{\sin (u)}{\sin (\lambda)}$ and $\psi_{a}$ is the
entry, associated with the node $a$, of the Perron-Frobenius eigenvector
of the adjacency matrix $G$.

These Boltzmann weights are represented by a local
face operator $X_{j}(u)$ in the Temperley-Lieb algebra
$\mathcal{T}(N,\, \lambda)$~\cite{BP01}:

\setlength{\unitlength}{14pt}
\begin{equation}
X_{j}(u)=
\vtop to 1.6\unitlength{}
\begin{picture}(2.5,1)(-.5,.9)
     \multiput(0,1)(1,-1){2}{\line(1,1){1}}
     \multiput(0,1)(1,1){2}{\line(1,-1){1}}
\put(1,1){\pp{}{u}}
\put(1,.3){\pp{}{\to}}
\multiput(0,-0.5)(0,0.25){13}{\pp{}{.}}
\multiput(1,-.5)(0,0.25){2}{\pp{}{.}}
\multiput(1,2.25)(0,0.25){2}{\pp{}{.}}
\multiput(2,-0.5)(0,0.25){13}{\pp{}{.}}
\put(0,-.6){\pp{t}{j\mi 1}}
\put(1,-.6){\pp{t}{j}}
\put(2,-.6){\pp{t}{j\plus 1}}
\end{picture}
\;=\; s(\lambda-u)\text{I}+
     s(u)e_{j}
\label{eq:X11}
\end{equation}
where $e_{j}=X_{j}(\lambda)$ is a Temperley-Lieb generator and
$j=1,\ldots,N$ labels the position in the lattice.

\subsection{Fusion Projector}
\label{sec:FusionProjector}
In turn, this model gives rise to a hierarchy of \emph{fused}  models
whose Boltzmann weights we are going to describe.


  We first define recursively the fusion operators $P^r_j$, for
  $r\in\< 1,g\>$ as follows:
\begin{equation}
\begin{array}{l}
P^1_j\;=\;P^2_j\;=\;I \\
{}\\
P^r_j\;=\;
\frac{1}{S_{r-1}}\;P^{r-1}_{j+1}\;X_j(-(r\mi2)\lambda)\:P^{r-1}_{j+1}\,,
\quad r\geq 3\, ,\end{array}
\label{eq:FusionOperator}
\end{equation}
where $S_{k}=s(k \lambda)$ and $j$ is suitably restricted~\cite{BP01}.
Thus, $P^r_j$ can be expressed as a function of $e_j,e_{j+1},\ldots,
e_{j+(r-3)}$. In particular,
\begin{equation}
P_{j}^{3}=\frac{1}{S_{2}}\;
\begin{picture}(2,1)(0,.9)
     \multiput(0,1)(1,-1){2}{\line(1,1){1}}
     \multiput(0,1)(1,1){2}{\line(1,-1){1}}
\put(1,1){\pp{}{-\lambda}}
\put(1,.3){\pp{}{\to}}
\end{picture}
= I - \frac{1}{S_{2}}\;
\begin{picture}(2,1)(0,.9)
     \multiput(0,1)(1,-1){2}{\line(1,1){1}}
     \multiput(0,1)(1,1){2}{\line(1,-1){1}}
\put(1,1){\pp{}{+\lambda}}
\put(1,.3){\pp{}{\to}}
\end{picture} \; .
	\label{eq:DefP}
\end{equation}

We shall represent the fusion operators diagrammatically as
\setlength{\unitlength}{5mm}
\begin{equation}
\raisebox{-1.75\unitlength}[1.75\unitlength][1.75\unitlength]{
\textB{1}{3.5}{0.5}{2}{}{P^r_j}
\textB{1.8}{3.5}{0.9}{2}{}{=}
\begin{picture}(6,3.5)
\multiput(0,2)(5,-1){2}{\line(1,1){1}}
\multiput(0,2)(5,1){2}{\line(1,-1){1}}
\multiput(1,1)(0,2){2}{\line(1,0){4}}
\multiput(0,0.5)(0,0.25){13}{\pp{}{.}}
\multiput(1,0.5)(0,0.25){2}{\pp{}{.}}
\multiput(1,3.25)(0,0.25){2}{\pp{}{.}}
\multiput(5,0.5)(0,0.25){2}{\pp{}{.}}
\multiput(5,3.25)(0,0.25){2}{\pp{}{.}}
\multiput(6,0.5)(0,0.25){13}{\pp{}{.}}
\put(-0.05,0){\pp{b}{j-1}}\put(1.1,0){\pp{b}{j}}
\put(5.3,0){\pp{br}{j+r-3}}\put(5.7,0){\pp{bl}{j+r-2}}
\put(6.4,1.7){\p{}{.}}\end{picture} }
\end{equation}

It is easy to show that this operator is in fact a projector.
Moreover,
\begin{gather}
     P_{j'}^{r'}P_{j}^{r}=P_{j}^{r}P_{j'}^{r'}=P_{j}^{r}\qquad
     \text{ for } 0\leq j'-j\leq r-r'.
     \label{eq:PP}\\
\setlength{\unitlength}{10pt}
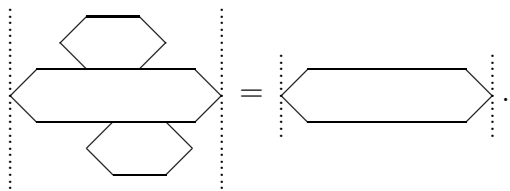
     \begin{picture}(8,3.5)(0,1.8)
\multiput(0,2)(7,-1){2}{\line(1,1){1}}
\multiput(0,2)(7,1){2}{\line(1,-1){1}}
\multiput(1,1)(0,2){2}{\line(1,0){6}}
\multiput(1.5,2)(1,-4){2}{
     	\begin{picture}(4,3.5)
\multiput(0,2)(3,-1){2}{\line(1,1){1}}
\multiput(0,2)(3,1){2}{\line(1,-1){1}}
\multiput(1,1)(0,2){2}{\line(1,0){2}}
	\end{picture}
	}
\multiput(0,-1.5)(0,0.25){28}{\pp{}{.}}
\multiput(8,-1.5)(0,0.25){28}{\pp{}{.}}
\end{picture}
\;=\;
     \begin{picture}(8,3.5)(0,1.8)
\multiput(0,2)(7,-1){2}{\line(1,1){1}}
\multiput(0,2)(7,1){2}{\line(1,-1){1}}
\multiput(1,1)(0,2){2}{\line(1,0){6}}
\multiput(0,.5)(0,0.25){13}{\pp{}{.}}
\multiput(8,.5)(0,0.25){13}{\pp{}{.}}
\end{picture}
\vtop to 3\unitlength { } \; .
\notag
\end{gather}
This implies that any operator expressible as a product of local face
operators and falling within the boundaries of a projector, acts as a
scalar on it:
\begin{gather}
    X_{j'}(u)P_{j}^{r}= P_{j}^{r}X_{j'}(u)=s(\lambda-u)P_{j}^{r}\qquad
     \text{ for } 0\leq j'-j\leq r-2.
     \label{eq:XP}\\
\setlength{\unitlength}{10pt}
     \begin{picture}(8,5)(0,1.8)
\multiput(0,2)(7,-1){2}{\line(1,1){1}}
\multiput(0,2)(7,1){2}{\line(1,-1){1}}
\multiput(1,1)(0,2){2}{\line(1,0){6}}
\put(2.1,4.7){
     	\begin{picture}(3,3)
     \multiput(0,0)(1,-1){2}{\line(1,1){1}}
     \multiput(0,0)(1,1){2}{\line(1,-1){1}}
\put(1,0){\pp{}{u}}
\put(1,-.7){\pp{}{\to}}
\end{picture}
	}
\multiput(0,.5)(0,0.25){22}{\pp{}{.}}
\multiput(8,.5)(0,0.25){22}{\pp{}{.}}
\end{picture}
\;=\;s(\lambda-u)\;\;
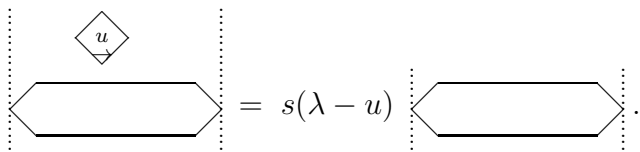
     \begin{picture}(8,3.5)(0,1.8)
\multiput(0,2)(7,-1){2}{\line(1,1){1}}
\multiput(0,2)(7,1){2}{\line(1,-1){1}}
\multiput(1,1)(0,2){2}{\line(1,0){6}}
\multiput(0,.5)(0,0.25){13}{\pp{}{.}}
\multiput(8,.5)(0,0.25){13}{\pp{}{.}}
\end{picture}
\vtop to 1.5\unitlength {} \; .
\notag
\end{gather}
A particularly important case is for $u=+\lambda$:  its local face
operator is a projector orthogonal to all the $P^{r}_{j}$.

  As $P^{r}_{j}$ is clearly translationally covariant (in its domain of
  definition) we can decompose it onto the spaces of paths with given
  end points: $P^{r}(a,b)$ is the fusion projector acting on paths from
  $a$ to $b$ in $r\mi 1$ steps.  Its rank is given by the \emph{fused
  adjacency matrix} entries:
\begin{equation}
     \text{Rank~}\left(P^{r}(a,b)\right)=F_{a\; b}^{r}
     \label{eq:RankP}
\end{equation}
also called \emph{basic intertwiners} and recursively defined by
the $\hat{s\ell}_{2}$  fusion rules:
\begin{equation}
     F^{1}=\text{I}, \;\;\;\; F^{2}=G,\;\;\;\; F^{r}=F^{r-1}F^{2}-F^{r-2},
     \text{ for } r=3,\ldots,g.
     \label{eq:F}
\end{equation}
This equation can be recast as
\begin{equation}
(\tilde{F}^r)^2\:=\: (I+\tilde{F}^{r-1})(I+\tilde{F}^{r+1})
\label{eq:Ft}
\end{equation}
where
\begin{equation}
\tilde{F}^r\:=\:F^{r-1}\:F^{r+1}\,.
\end{equation}

The $+1$ eigenvectors of $P^{r}(a,b)$ are thus indexed by
an integer $\gamma\in\< 1,F_{a\; b}^{r}\>$ refered to as the \emph{bond
variable}. We denote them by $\vec U^{r}_{\gamma}(a,b)$ and call them
\emph{fusion vectors}.

\subsection{Fused face operators}
\label{sec:FusedFace}
These projectors allow us to define the $(p,q)$-fused face operator
defined as the product of $q$ rows of $p$ local face
operators with a shift of the spectral parameter by $\pm \lambda$ from
one face to the next:
\setlength{\unitlength}{14pt}
\begin{equation}
\textB{3.5}{0}{0.5}{2}{}{X^{pq}_{j}(u)=}
\begin{picture}(5,6)
     \multiput(0,2)(2,-2){2}{\line(1,1){3}}
     \multiput(0,2)(3,3){2}{\line(1,-1){2}}
\put(2.5,2.5){\pp{}{X^{pq}(u)}}
\put(2,.3){\pp{}{\to}}
\multiput(0,-0.5)(0,0.25){25}{\pp{}{.}}
\multiput(2,-.5)(0,0.25){2}{\pp{}{.}}
\multiput(3,5.25)(0,0.25){2}{\pp{}{.}}
\multiput(5,-0.5)(0,0.25){25}{\pp{}{.}}
\put(0,-.6){\pp{t}{j-1}}
\put(2,-.6){\pp{t}{j+ q\mi 1}}
\put(3,5.75){\pp{b}{j+ p\mi 1}}
\put(5,-.6){\pp{t}{j+ p+ q\mi 2}}\end{picture}
\textB{2}{0}{1}{2}{}{=}
\begin{picture}(7,6)(-2,0)
     \multiput(0,2)(.5,-.5){2}{\line(1,1){3}}
     \multiput(2,0)(-.5,.5){2}{\line(1,1){3}}
     \multiput(0,2)(.5,.5){2}{\line(1,-1){2}}
     \multiput(3,5)(-.5,-.5){2}{\line(1,-1){2}}
\put(4.5,3){\pp{}{u}}
\put(2.,5.5){\vector(1,-1){1}}
\put(2.,5.25){\pp{br}{u+(q\mi 1)\lambda}}
\put(.5,3.75){\vector(0,-1){1.75}}
\put(-.5,4.5){\pp{}{u+(q\mi p)\lambda}}
\put(3,-1){\vector(-2,3){1}}
\put(3.1,-1){\pp{tl}{u-(p\mi 1)\lambda}}
\put(4.2,3.3){\vector(-1,1){.85}}
\put(.8,2.3){\vector(1,1){1.85}}
\put(1.7,0.8){\vector(-1,1){.85}}
\put(2.3,.8){\vector(1,1){1.85}}
\multiput(0,-0.5)(0,0.25){16}{\pp{}{.}}
\multiput(2,-.5)(0,0.25){2}{\pp{}{.}}
\multiput(3,5.25)(0,0.25){2}{\pp{}{.}}
\multiput(5,-.5)(0,0.25){25}{\pp{}{.}}
\put(0,0){\pp{lb}{\diagup}}
\put(0,0){\pp{lt}{\diagdown}}
\multiput(.5,.5)(0,-1){2}{\line(1,0){1}}
\put(1,0){\pp{}{P_{j}^{q\plus 1}}}
\put(2,0){\pp{tr}{\diagup}}
\put(2,0){
\put(0,0){\pp{lt}{\diagdown}}
\multiput(.5,.5)(0,-1){2}{\line(1,0){2}}
\put(1.5,0){\pp{}{P_{\!j\plus q}^{p\plus 1}}}
\put(3,0){\pp{tr}{\diagup}}
\put(3,0){\pp{rb}{\diagdown}}
}
\end{picture}
\textB{2}{0}{1}{2}{}{.}
\vtop to 1.5\unitlength {}
            \label{eq:Xpq}
\end{equation}
The position of the projectors and spectral parameters can be
  altered by \emph{pushing-through}:
\begin{equation}
\textB{2.25}{0}{0.5}{2}{}{X^{pq}_{j}(u)=}
\begin{picture}(6,5.5)(-1,0)
     \multiput(0,2)(.5,-.5){2}{\line(1,1){3}}
     \multiput(2,0)(-.5,.5){2}{\line(1,1){3}}
     \multiput(0,2)(.5,.5){2}{\line(1,-1){2}}
     \multiput(3,5)(-.5,-.5){2}{\line(1,-1){2}}
\put(3,4.5){\pp{}{u}}
\put(.5,3.5){\vector(0,-1){1.5}}
\put(0,3.6){\pp{b}{u-(p\mi 1)\lambda}}
\put(1,-1){\vector(2,3){1}}
\put(1,-1){\pp{t}{u+(q\mi p)\lambda}}
\put(3.3,4.2){\vector(1,-1){.85}}
\put(.8,2.3){\vector(1,1){1.85}}
\put(.8,1.7){\vector(1,-1){.85}}
\put(2.3,.8){\vector(1,1){1.85}}
\multiput(0,-0.5)(0,0.25){16}{\pp{}{.}}
\multiput(2,-.5)(0,0.25){2}{\pp{}{.}}
\multiput(3,5.25)(0,0.25){4}{\pp{}{.}}
\multiput(5,-.5)(0,0.25){27}{\pp{}{.}}
\put(3,5){
\put(0,0){\pp{lb}{\diagup}}
\multiput(.5,.5)(0,-1){2}{\line(1,0){1}}
\put(1,0){\pp{}{P_{\!j\plus p}^{q\plus 1}}}
\put(2,0){\pp{tr}{\diagup}}
\put(2,0){\pp{rb}{\diagdown}}
}
\put(2,0){
\put(0,0){\pp{lt}{\diagdown}}
\multiput(.5,.5)(0,-1){2}{\line(1,0){2}}
\put(1.5,0){\pp{}{P_{\!j\plus q}^{p\plus 1}}}
\put(3,0){\pp{tr}{\diagup}}
\put(3,0){\pp{rb}{\diagdown}}
}
\end{picture}
\textB{2}{0}{1}{2}{}{=}
\begin{picture}(4.5,5.5)(0,0)
     \multiput(0,2)(.5,-.5){2}{\line(1,1){3}}
     \multiput(2,0)(-.5,.5){2}{\line(1,1){3}}
     \multiput(0,2)(.5,.5){2}{\line(1,-1){2}}
     \multiput(3,5)(-.5,-.5){2}{\line(1,-1){2}}
\put(.5,2){\pp{}{u}}
\put(3,-1){\vector(-2,3){1}}
\put(3.1,-1){\pp{t}{u+(q\mi 1)\lambda}}
\put(3.3,4.2){\vector(1,-1){.85}}
\put(2.7,4.2){\vector(-1,-1){1.85}}
\put(.8,1.7){\vector(1,-1){.85}}
\put(4.2,2.7){\vector(-1,-1){1.85}}
\multiput(0,-0.5)(0,0.25){27}{\pp{}{.}}
\multiput(2,-.5)(0,0.25){2}{\pp{}{.}}
\multiput(3,5.25)(0,0.25){4}{\pp{}{.}}
\multiput(5,-.5)(0,0.25){27}{\pp{}{.}}
\put(3,5){
\put(0,0){\pp{lb}{\diagup}}
\multiput(.5,.5)(0,-1){2}{\line(1,0){1}}
\put(1,0){\pp{}{P_{\!j\plus p}^{q\plus 1}}}
\put(2,0){\pp{tr}{\diagup}}
\put(2,0){\pp{rb}{\diagdown}}
}
\put(0,5){
\put(0,0){\pp{lb}{\diagup}}
\put(0,0){\pp{lt}{\diagdown}}
\multiput(.5,.5)(0,-1){2}{\line(1,0){2}}
\put(1.5,0){\pp{}{P_{j}^{p\plus 1}}}
\put(3,0){\pp{rb}{\diagdown}}
}
\end{picture}
\textB{2.5}{0}{1.5}{2}{}{=}
\begin{picture}(5,5.5)(0,0)
     \multiput(0,2)(.5,-.5){2}{\line(1,1){3}}
     \multiput(2,0)(-.5,.5){2}{\line(1,1){3}}
     \multiput(0,2)(.5,.5){2}{\line(1,-1){2}}
     \multiput(3,5)(-.5,-.5){2}{\line(1,-1){2}}
\put(2,.5){\pp{}{u}}
\put(4.2,3.3){\vector(-1,1){.85}}
\put(2.7,4.2){\vector(-1,-1){1.85}}
\put(1.7,0.8){\vector(-1,1){.85}}
\put(4.2,2.7){\vector(-1,-1){1.85}}
\multiput(0,-1)(0,0.25){29}{\pp{}{.}}
\multiput(2,-1)(0,0.25){4}{\pp{}{.}}
\multiput(3,5.25)(0,0.25){4}{\pp{}{.}}
\multiput(5,-1)(0,0.25){29}{\pp{}{.}}
\put(0,0){
\put(0,0){\pp{lb}{\diagup}}
\put(0,0){\pp{lt}{\diagdown}}
\multiput(.5,.5)(0,-1){2}{\line(1,0){1}}
\put(1,0){\pp{}{P_{j}^{q\plus 1}}}
\put(2,0){\pp{tr}{\diagup}}
}
\put(0,5){
\put(0,0){\pp{lb}{\diagup}}
\put(0,0){\pp{lt}{\diagdown}}
\multiput(.5,.5)(0,-1){2}{\line(1,0){2}}
\put(1.5,0){\pp{}{P_{j}^{p\plus 1}}}
\put(3,0){\pp{rb}{\diagdown}}
}
\end{picture}\,
\textB{2}{0}{1}{2}{}{.}
            \label{eq:PushThru}
\end{equation}
These properties imply several others, namely the \emph{Transposition
Symmetry},
\begin{equation}
     X^{pq}_{j}(u)^{T}=X^{qp}_{j}(u+(q\mi p)\lambda),
     \label{eq:Transpose}
\end{equation}
the \emph{Generalized
Yang-Baxter Equation} (GYBE),
\newsavebox{\Xpq}
\newsavebox{\Xqp}
\newsavebox{\Xqq}
\begin{equation}
\setlength{\unitlength}{17pt}
    \savebox{\Xpq}(3,3){
    \begin{picture}(3,3)(0,2)
     \multiput(0,2)(2,-2){2}{\line(1,1){1}}
     \multiput(0,2)(1,1){2}{\line(1,-1){2}}
\put(1.5,1.5){\pp{}{X^{pq'}(u)}}
\put(2,.3){\pp{}{\to}}
\end{picture}}
    \savebox{\Xqp}(4,4){
    \begin{picture}(4,4)(0,1)
     \multiput(0,1)(1,-1){2}{\line(1,1){3}}
     \multiput(0,1)(3,3){2}{\line(1,-1){1}}
\put(2,2){\pp{}{X^{qp}(v)}}
\put(1,.3){\pp{}{\to}}
\end{picture}}
    \savebox{\Xqq}(4,5){
    \begin{picture}(4,5)(0,2)
     \multiput(0,2)(2,-2){2}{\line(1,1){3}}
     \multiput(0,2)(3,3){2}{\line(1,-1){2}}
\put(2.5,2.5){\pp{}{X^{qq'}(u+v)}}
\put(2,.3){\pp{}{\to}}
\end{picture}}
\begin{picture}(8,8.5)(0,-.5)
\multiput(0,-.5)(0,0.25){33}{\pp{l}{.}}
\multiput(2,-.5)(0,0.25){2}{\pp{l}{.}}
\multiput(3,-.5)(0,0.25){6}{\pp{l}{.}}
\multiput(3,7.25)(0,0.25){2}{\pp{l}{.}}
\multiput(4,6.25)(0,0.25){6}{\pp{l}{.}}
\multiput(6,-.5)(0,0.25){33}{\pp{l}{.}}
     \put(0,4){\usebox{\Xqp}}
     \put(1,3){\usebox{\Xqq}}
     \put(0,2){\usebox{\Xpq}}
     \put(7,3){=}
\end{picture}
\begin{picture}(6,8.5)(0,-.5)
\multiput(0,-.5)(0,0.25){33}{\pp{l}{.}}
\multiput(2,-.5)(0,0.25){6}{\pp{l}{.}}
\multiput(3,-.5)(0,0.25){2}{\pp{l}{.}}
\multiput(3,6.25)(0,0.25){6}{\pp{l}{.}}
\multiput(4,7.25)(0,0.25){2}{\pp{l}{.}}
\multiput(6,-.5)(0,0.25){33}{\pp{l}{.}}
     \put(3,6){\usebox{\Xpq}}
     \put(0,3){\usebox{\Xqq}}
     \put(2,1){\usebox{\Xqp}}
\end{picture}
		\label{eq:GYBE}
\end{equation}
the \emph{Inversion Relation},
\begin{gather}
\setlength{\unitlength}{17pt}
   X^{pq}_{j}(u)
     X^{qp}_{j}(-u)
\;=\;
\begin{picture}(7.8,2)(-.5,1)
     \multiput(0,1)(1,-1){2}{\line(1,1){2}}
     \multiput(0,1)(2,2){2}{\line(1,-1){1}}
\put(1.5,1.5){\pp{}{X^{pq}(u)}}
\put(.3,1){\pp{}{\downarrow}}
   \put(3,2){\begin{picture}(4,4)(0,1)
     \multiput(0,1)(1,1){2}{\line(1,-1){2}}
     \multiput(0,1)(2,-2){2}{\line(1,1){1}}
\put(1.5,.5){\pp{}{X^{qp}(-u)}}
\put(.3,1){\pp{}{\downarrow}}
\end{picture}}
\multiput(-.5,0)(0.25,0){30}{\pp{}{.}}
\multiput(-.5,3)(0.25,0){30}{\pp{}{.}}
\multiput(-.5,1)(0.25,0){2}{\pp{}{.}}
\multiput(6.25,1)(0.25,0){2}{\pp{}{.}}
\end{picture}
\notag\\[1.5\unitlength]
=\;s_{1}^{p\,q}(u)\,s_{1}^{p\,q}(-u)\:P_{j}^{q+1}P_{j+q}^{p+1}\,,
		\label{eq:InversionRelation}
\end{gather}
where $s^{p\,q}_{i}(u)=
\prod_{j=0}^{p-1}\prod_{k=0}^{q-1}s(u+(i\mi j\plus k)\lambda)$,
and the \emph{Abelian Property},
\begin{gather}
\setlength{\unitlength}{17pt}
   X^{pq}_{j}(u+(p\mi 1)\lambda)
     X^{qp}_{j}(v+(q\mi 1)\lambda)
=  X^{pq}_{j}(v+(p\mi 1)\lambda)
     X^{qp}_{j}(u+(q\mi 1)\lambda)
\,.
		\label{eq:Abelian}
		\\ \notag
\end{gather}

These operators, contracted against the fusion vectors, yield the
$(p,q)$-fused Boltzmann weights. They depend not only on the spins on
the four corners but also on bond variables on the edges:
\begin{gather}
    \kern-3em
     \setlength{\unitlength}{.8cm}
      W^{pq}\W{
     d&\gamma& c\\
     \delta&&\beta\\
     a&\alpha&b
     }{|u}=
     \begin{picture}(2,0)(0,.85)
\put(0.5,0.5){\framebox(1,1){$
u
$}}
\put(.5,.52){\pp{bl}{\sss \searrow}}
\put(0.6,0.62){\pp{bl}{\sss pq}}
\put(0.45,0.45){\pp{tr}{a}}
\put(1,.5){\pp{}{\alpha}}
\put(1.55,0.45){\pp{tl}{b}}
\put(1.5,1){\pp{}{\beta}}
\put(1.55,1.55){\pp{bl}{c}}
\put(1,1.5){\pp{}{\gamma}}
\put(0.45,1.55){\pp{br}{d}}
\put(0.5,1){\pp{}{\delta}}
\end{picture}
=
     \frac{1}{s^{p\, q\mi 1}_{0}(u)}
     \setlength{\unitlength}{.7cm}
\vtop to 2\unitlength{}
\begin{picture}(6,2.25)(-2.5,-.25)
\put(0,-1){\framebox(3,2){\p{}{X^{pq}(u)}}}
\put(.1,-.9){\pp{bl}{\searrow}}
\put(-.25,-1.25){\pp{}{a}}
\put(1.5,-1.4){\pp{}{\vec U^{p-1}_{\alpha}(a,b)}}
\put(3.25,-1.25){\pp{}{b}}
\put(3.2,0){\pp{l}{{\vec U^{q-1}_{\beta}(a,b)}^{\dag}}}
\put(3.25,1.25){\pp{}{c}}
\put(1.5,1.4){\pp{}{{{\vec U^{p-1}_{\gamma}(d,c)}}^{\dag}}}
\put(-.25,1.25){\pp{}{d}}
\put(-.2,0){\pp{r}{\vec U^{q-1}_{\delta}(a,b)}}
\end{picture}\notag \\
            \label{eq:Wpq}
\end{gather}
where the normalization function $s^{p\, q\mi1}_{0}(u)$
eliminates some scalar factors common to all the spin configurations
which appear in the process of fusion.
In the $A_{L}$ case, the bond variables are trivial, that is,
$\alpha,\beta,\gamma,\delta=1$.

The fused Boltzmann weights satisfy a \emph{Diagonal Reflection}
\begin{gather}
     W^{pq}\W{
     d&\gamma& c\\
     \delta&&\beta\\
     a&\alpha&b
     }{|u}
=
\frac{s^{q\, p\mi1}_{q\mi p}(u)}{s^{p\, q-1}_{0}(u)}
     \;
     W^{qp}\W{
     d&\delta& a\\
     \gamma&&\alpha\\
     c&\beta&b
     }{|u+(q\mi p)\lambda},
     \label{eq:WSymmetry}
\end{gather}
and
\emph{Crossing Symmetry}:
\begin{gather}
     W^{pq}\W{
     d&\gamma& c\\
     \delta&&\beta\\
     a&\alpha&b
     }{|u}
=\sqrt{\frac{\psi_{a}\psi_{c}}{\psi_{b}\psi_{d}}}
      \;
    \frac{s^{q\, p\mi 1}_{0}(\lambda\mi u)}
     {s^{p\, q\mi 1}_{0}(u)}
     \;
     W^{qp}\W{
     a&\delta& d\\
     \alpha&&\gamma\\
     b&\beta&c
     }{|\lambda\mi u}.
     \label{eq:CrossingSymmetry}
\end{gather}

\section{Transfer matrix}
\label{sec:Transfer}\setcounter{equation}{0}
Given the hierarchy of fused Boltzmann weights, we build transfer
matrices for different fusion levels and boundary
conditions: on the torus, and on the cylinder, with or without seams.

\subsection{Seams}
\label{sec:Seam}
Simple seams are modified faces.  They come in three different types,
$r$, $s$ and $\zeta$-type.  A label $(r,s,\zeta)\in A_{g-2}\times
A_{g-1}\times \Gamma$, where $\Gamma$ is the symmetry algebra of the
graph $G$, encodes a triple seam involving three modified faces.  The
symmetry $\zeta$ is taken as the identity when omitted.

We first define $W^{q}_{(r,1)}$, the $r$-type seam for the
$(p,q)$-fused model.  It is a usual $(r-1,q)$-fused face (it does not
depend on the horizontal fusion level $p$) with a boundary field
$\xi$ acting as a shift in the spectral parameter, and another choice
for the removal of the common scalar factors:
\begin{equation}
    \kern-3em
   W^{q}_{(r,1)}\W{d&\gamma&c\\ \delta&&\beta\\ a&\alpha &
   b}{|u,\xi}=\!\!
   \setlength{\unitlength}{10mm}
\begin{picture}(1.5,0)(.2,.85)
\multiput(0.5,0.5)(1,0){2}{\line(0,1){1}}
\multiput(0.5,0.5)(0,1){2}{\line(1,0){1}}
\put(.5,.52){\pp{bl}{\sss \searrow}}
\put(0.55,0.57){\pp{bl}{\sss q}}
\put(0.45,0.45){\pp{tr}{a}}
\put(1,.5){\pp{}{\alpha}}
\put(1.55,0.45){\pp{tl}{b}}
\put(1.5,1){\pp{}{\beta}}
\put(1.55,1.55){\pp{bl}{c}}
\put(1,1.5){\pp{}{\gamma}}
\put(0.45,1.55){\pp{br}{d}}
\put(0.5,1){\pp{}{\delta}}
\put(1,1){\pp{}{{}^{r}\!(u,\xi)}}
\end{picture}
		= \,
   \frac{s_{0}^{r\mi 1\, q\mi 1}(u+\xi)}{s_{-1}^{r\mi 2\, q}(u+\xi)}
   \; W^{(r\mi 1)\, q}\W{
     d&\gamma& c\\
     \delta&&\beta\\
     a&\alpha&b
     }{|u+\xi}.
     \label{eq:rSeam}
\end{equation}

An $s$-type seam is the normalized \emph{braid limit} of an $r$-type seam, it
does not depend on any spectral parameter:
\begin{equation}
    \kern-3em
   W^{q}_{(1,s)}\W{d&\gamma&c\\ \delta&&\beta\\ a&\alpha &
   b}{.}=\!
   \setlength{\unitlength}{8mm}
\begin{picture}(2,0)(0,.85)
\multiput(0.5,0.5)(1,0){2}{\line(0,1){1}}
\multiput(0.5,0.5)(0,1){2}{\line(1,0){1}}
\put(.5,.52){\pp{bl}{\sss \searrow}}
\put(0.55,0.57){\pp{bl}{\sss q}}
\put(0.45,0.45){\pp{tr}{a}}
\put(1,.5){\pp{}{\alpha}}
\put(1.55,0.45){\pp{tl}{b}}
\put(1.5,1){\pp{}{\beta}}
\put(1.55,1.55){\pp{bl}{c}}
\put(1,1.5){\pp{}{\gamma}}
\put(0.45,1.55){\pp{br}{d}}
\put(0.5,1){\pp{}{\delta}}
\put(1,1){\pp{}{(1,s)}}
\end{picture}
			\! = \;
   \lim_{\xi\to i \infty}\;\frac{e^{-i\frac{(g+1)(s-1)q}{2}\lambda}}
   {s_{0}^{1q}(u+\xi)}\;
   W^{q}_{(s,1)}\W{d&\gamma&c\\ \delta&&\beta\\ a&\alpha & b}{|u,\xi}.
     \label{eq:sSeam}
\end{equation}

The automorphisms $\zeta\in\Gamma$ of the adjacency matrix, satisfying
$G_{a,b}=G_{\zeta(a),\zeta(b)}$, leave the face weights invariant
\begin{equation}
    \kern-3em
W^{pq}\W{
d&\gamma&c\\
\delta&&\beta\\
a&\alpha & b}{|u}\;=\;
\setlength{\unitlength}{8mm}
\begin{picture}(2,0)(0,.85)
\multiput(0.5,0.5)(1,0){2}{\line(0,1){1}}
\multiput(0.5,0.5)(0,1){2}{\line(1,0){1}}
\put(.5,.52){\pp{bl}{\sss \searrow}}
\put(0.6,0.62){\pp{bl}{\sss pq}}
\put(0.45,0.45){\pp{tr}{a}}
\put(1,.5){\pp{}{\alpha}}
\put(1.55,0.45){\pp{tl}{b}}
\put(1.5,1){\pp{}{\beta}}
\put(1.55,1.55){\pp{bl}{c}}
\put(1,1.5){\pp{}{\gamma}}
\put(0.45,1.55){\pp{br}{d}}
\put(0.5,1){\pp{}{\delta}}
\put(1,1){\pp{}{u}}
\end{picture}
			\; = \;
\begin{picture}(2,0)(0,.85)
\multiput(0.5,0.5)(1,0){2}{\line(0,1){1}}
\multiput(0.5,0.5)(0,1){2}{\line(1,0){1}}
\put(.5,.52){\pp{bl}{\sss \searrow}}
\put(0.6,0.62){\pp{bl}{\sss pq}}
\put(0.45,0.45){\pp{tr}{\zeta(a)}}
\put(1,.5){\pp{}{\alpha}}
\put(1.55,0.45){\pp{tl}{\zeta(b)}}
\put(1.5,1){\pp{}{\beta}}
\put(1.55,1.55){\pp{bl}{\zeta(c)}}
\put(1,1.5){\pp{}{\gamma}}
\put(0.45,1.55){\pp{br}{\zeta(d)}}
\put(0.5,1){\pp{}{\delta}}
\put(1,1){\pp{}{u}}
\end{picture}
			\; = \;
W^{pq}\W{
\zeta(d)&\gamma&\zeta(c)\\
\delta        &&\beta\\
\zeta(a)&\alpha&\zeta(b)}{|u}
\end{equation}
and act through the special seam~\cite{CKP}
\begin{equation}
    \kern-3em
\setlength{\unitlength}{8mm}
       {\ds W^{q}_{(1,1,\zeta)}}
       \W{d&c\\ \alpha&\beta\\ a & b}{.}\;=
       \delta_{b\,\zeta(a)}\delta_{c\,\zeta(d)}
\begin{picture}(2,0)(0,.85)
\multiput(0.5,0.5)(1,0){2}{\line(0,1){1}}
\multiput(0.5,0.5)(0,1){2}{\line(1,0){1}}
\put(0.5,0.52){\pp{bl}{\sss\searrow}}
\put(0.55,0.57){\pp{bl}{\sss q}}
\put(0.45,0.45){\pp{tr}{a}}\put(1.55,0.45){\pp{tl}{\zeta(a)}}
\put(1.55,1.55){\pp{bl}{\zeta(d)}}\put(0.45,1.55){\pp{br}{d}}
\put(1.5,1){\pp{}{\beta}}
\put(0.5,1){\pp{}{\alpha}}
\put(1,1){\pp{}{\zeta}}
\end{picture}
\;=\;
\cases{1, &F_{a\,d}^{q+1}\neq 0,\ \alpha=\beta\\ & b=\zeta(a),\ c=\zeta(d),\cr
        0, & \text{otherwise.}}
       \label{eq:seamWzeta}
\end{equation}
Notice that the $(r,s,\zeta)=(1,1,1)$ seam, where $\zeta=1$ denotes
the identity automorphism, is the empty seam
\begin{equation}
    W^{q}_{(1,1,1)}\W{d&c\\ \alpha&\beta\\ a & b}{.}\;=
\;\delta_{ab}\;\delta_{cd}\;\delta_{\alpha\beta}\;F_{b\,c}^{q+1}.
       \label{eq:W111}
\end{equation}
The push-through property is also trivially verified for a $\zeta$-type
seam.

The label $s$ appearing in a $(1,s)$-seam is an integer in $A_{g-1}$.
In~\cite{CMOP2}, we define an $(1,a)$-seam with $a\in G$ which reduces
to the definition given here for $G$ of $A$ type but which extends it
for the $D_{\text{even}}$, $E_{6}$ and  $E_{8}$ graphs.

\subsection{Torus transfer matrix}
\label{sec:TorusTransfer}
The transfer matrix for the $(p,q)$-fused model with an
$(r,s,\zeta)$-seam, on the $N$ faces torus on the square lattice is
given, in the basis of the cyclic paths in $N$ steps plus the seam, with bond
variables between adjacent spins, by the product of the corresponding
Boltzmann weights:
The entries of the transfer matrix with an $(r,s,\zeta)$ seam are
given by
\begin{gather}
    \kern-3em
\rule{0pt}{24pt}\< \vec{a},\vec{\alpha}|\;
\vec{T}_{(r,s,\zeta)}^{pq}(u,\xi)\;
        |\vec{b},\vec{\beta}\>=
\setlength{\unitlength}{12mm}
\vtop to .7\unitlength {}
\begin{picture}(7.2,.8)(.4,.9)
      \multiput(0.5,0.5)(0,1){2}{\line(1,0){7}}
\multiput(0.5,0.5)(1,0){2}{\line(0,1){1}}
\multiput(3.5,0.5)(1,0){5}{\line(0,1){1}}
\put(0.5,0.52){\pp{bl}{\sss\searrow}}
\put(0,0){\put(0.5,0.55){\pp{bl}{\sss pq}}}
\multiput(3,0)(1,0){4}{
\put(0.5,0.52){\pp{bl}{\sss\searrow}}
}
\put(2.5,1){\pp{}{\cdots}}
\put(3,0){\put(0.5,0.55){\pp{bl}{\sss pq}}}
\multiput(4,0)(1,0){3}{\put(0.55,0.57){\pp{bl}{\sss q}}}
\put(0.5,0.35){\pp{c}{a_{1}}}
\put(1,0.5){\pp{c}{\alpha_{1}}}
\put(1.5,0.35){\pp{c}{a_{2}}}
\put(3.5,0.35){\pp{c}{a_{N}}}
\put(4,0.5){\pp{c}{\alpha_{N}}}
\put(4.5,0.35){\pp{c}{a_{N\plus1}}}
\put(5,0.5){\pp{c}{\,\alpha_{N\plus1}}}
\put(5.5,0.35){\pp{c}{\,a_{N\plus2}}}
\put(6,0.5){\pp{c}{\alpha_{N\plus2}}}
\put(6.5,0.35){\pp{c}{a_{N\plus3}}}
\put(7.5,0.35){\pp{c}{a_{1}}}
\put(0.5,1.7){\pp{c}{b_{1}}}
\put(1,1.5){\pp{c}{\beta_{1}}}
\put(1.5,1.7){\pp{c}{b_{2}}}
\put(3.5,1.7){\pp{c}{b_{N}}}
\put(4,1.5){\pp{c}{\beta_{N}}}
\put(4.5,1.7){\pp{c}{b_{N\plus1}}}
\put(5,1.5){\pp{c}{\beta_{N\plus1}}}
\put(5.5,1.7){\pp{c}{b_{N\plus2}}}
\put(6,1.5){\pp{c}{\beta_{N\plus2}}}
\put(6.5,1.7){\pp{c}{b_{N\plus3}}}
\put(7.5,1.7){\pp{c}{b_{1}}}
\put(1,1){\pp{}{u}}
\put(4,1){\pp{}{u}}
\put(5,1){\pp{}{{}^{r}\!(u,\xi)}}
\put(6,1){\pp{}{{(1,s)}}}
\put(7,1){\pp{}{\zeta}}
\put(2.1,0.8){\p{}{}}\end{picture}
=
        \notag \\[16pt]
     \sum_{\vec\gamma}\prod_{i=1}^{N}W^{pq}\W{b_{i}&\beta_{i}&b_{i+1}\\
     \gamma_{i}&&\gamma_{i\plus 1}\\
     a_{i}&\alpha_{i}&a_{i+1}}{|u}
     W^{q}_{(r,1)}\W{b_{N}&\beta_{N}&b_{N+1}\\
     \gamma_{N}&&\gamma_{N\plus 1}\\
     a_{N}&\alpha_{N}&a_{N+1}}{|u,\xi}\times\qquad\notag\\ \qquad
     W^{q}_{(1,s)}\W{b_{N\plus 1}&\beta_{N\plus 1}&b_{N\plus 2}\\
     \gamma_{N\plus 1}&&\gamma_{N\plus 2}\\
     a_{N\plus 1}&\alpha_{N\plus 1}&a_{N\plus 2}}{.}
     W^{q}_{(1,1,\zeta)}\W{b_{N\plus 2}&b_{1}\\
     \gamma_{N\plus 2}&\gamma_{1}\\
     a_{N\plus 2}&a_{1}}{.}
         \label{eq:TpqTorus}
\end{gather}
where the sum is over all possible vertical bond variables.  The usual
periodic boundary condition is obtained for $(r,s,\zeta)=(1,1,1)$.
The definition can be generalised to accommodate an arbitrary number of
seams.  Because the seam faces are modified bulk faces, they satisfy
the GYBE, so they can be moved around freely with respect to the bulk
faces, the spectrum of the corresponding transfer matrices remains
unchanged.  However, in the $D_{2k}$ cases, when there are several
seams, their order can not be exchanged because the fusion algebra of
defect lines is non commutative~\cite{PZ01, CMOP2}.

\subsection{Boundary weights}
\label{sec:BoundaryWeigths}
The boundary weights are labelled by $(r,a)$ with $r\in A_{g-2}$ a
fusion level and $a\in G$ a node of the graph.

In the $A_{L}$ case, all $(r,s)$ boundary weights are obtained from
the action of an $(r,s)$-seam on the vacuum boundary
weight~\cite{BP96} and we construct in~\cite{CMOP2} an
$(r,a)$-seam with $a\in G$ so that it is also the case for the
$D_{\text{even}}$, $E_{6}$ and
$E_{8}$ graphs.  Nevertheless, in all cases, the $(1,a)$ boundary weights,
for two $q$-adjacent nodes of $G$, $c$ and $a$ (i.e. $F^{q+1}_{a\,
c}\neq 0$) are given explicitly by
\begin{gather}
     B^{q}_{(1,a)}\B{&&a\\
	&\gamma\\
        c\\
	&\alpha\\
		&&a}{.}=
\setlength{\unitlength}{8mm}
\begin{picture}(1.4,1)(-.25,.9)
\put(0,1){\line(1,-1){1}}
\put(0,1){\line(1,1){1}}
\multiput(1,0)(0,.205){10}{\line(0,1){.15}}
\put(.6,1){\pp{}{(1,a)}}
\put(1.1,2){\pp{l}{a}}
\put(.5,1.5){\pp{}{\gamma}}
\put(0,1){\pp{r}{c\,}}
\put(.5,.5){\pp{}{\alpha}}
\put(1.1,0){\pp{l}{a}}
\end{picture}
= {\frac{\psi_{c}^{\half}}{\psi_{a}^{\half}}} \;
\vec U^{q+1}_{\gamma}(c,a)^{\dag}\vec U^{q+1}_{\alpha}(c,a)=
{\frac{\psi_{c}^{\half}}{\psi_{a}^{\half}}}\;\delta_{\gamma\alpha}\;.
		\label{eq:B1a}
\end{gather}
   The vacuum boundary condition usually\footnote{When
extra structure is imposed, like in the superconformal
case~\cite{RP01}, the vacuum of the problem can be more complicated.}
corresponds to $(1,a)=(1,1)$.  The full $(r,a)$ boundary weights are
then given by the action of an $r$-type seam onto the $(1,a)$-boundary
weight.  The double row seam is given by two regular $r$-seams sharing
the same boundary field $\xi$, placed on top of one another,
with the same spectral parameters as bulk faces appearing in the
double row transfer matrix defined below in~\eqref{eq:TpqCylinder}:
\begin{gather}
     B^{q}_{(r,a)}\B{&&d&\; \delta\\
	&\gamma\\
        c\\
	&\alpha\\
		&&b&\; \beta}{|u,\xi}=
\setlength{\unitlength}{9mm}
\begin{picture}(1.8,1.2)(-.25,.9)
\put(1,2){\line(1,0){.5}}
\put(0,1){\line(1,-1){1}}
\put(0,1){\line(1,1){1}}
\put(1,0){\line(1,0){.5}}
\multiput(1.5,0)(0,.205){10}{\line(0,1){.15}}
\put(1,1.2){\pp{}{(r,a)}}
\put(1,.8){\pp{}{(u,\xi)}}
\put(1.5,2){\pp{l}{\,a}}
\put(1.25,2){\pp{}{\delta}}
\put(1,2.1){\pp{b}{d}}
\put(.5,1.5){\pp{}{\gamma}}
\put(0,1){\pp{r}{c\;}}
\put(.5,.5){\pp{}{\alpha}}
\put(1,-.1){\pp{t}{b}}
\put(1.25,0){\pp{}{\beta}}
\put(1.5,0){\pp{l}{\,a}}
\end{picture}
\;=
\;
\setlength{\unitlength}{9mm}
\begin{picture}(3.4,1)(-2.8,.9)
\put(-2.5,0){
\put(0,0){\framebox(2.5,2){}}
\put(0,1){\line(1,0){2.5}}
\multiput(0,0)(0,1){2}{
\put(0.,.05){\pp{bl}{\sss\searrow}\pp{bl}{\,\sss q}}
}}
\put(0,1){\line(1,-1){1}}
\put(0,1){\line(1,1){1}}
\multiput(1,0)(0,.205){10}{\line(0,1){.15}}
\multiput(.2,2)(.2,0){4}{\pp{}{.}}
\multiput(.2,0)(.2,0){4}{\pp{}{.}}
\put(-1.25,1.6){\pp{}{{}^{r}\!(\mu\mi u-{\sss (q\mi 1)}\lambda,\, \xi)}}
\put(-1.25,.5){\pp{}{{}^{r}\!(u,\xi)}}
\put(.6,1){\pp{}{(1,a)\,}}
\put(1.1,2){\pp{l}{a}}
\put(-1.25,2){\pp{r}{\delta}}
\put(-2.5,2.1){\pp{b}{d}}
\put(-2.5,1.5){\pp{}{\gamma\,}}
\put(-2.5,1){\pp{r}{c\;\;{}}}
\put(-2.5,.5){\pp{}{\alpha}}
\put(-2.5,-.1){\pp{t}{b}}
\put(-1.25,0){\pp{r}{\beta}}
\put(1.1,0){\pp{l}{a}}
\end{picture}
	\label{eq:Bra}
\end{gather}
and the left boundary weights are simply equal to the right boundary
weights.

These boundary weights satisfy boundary versions of the equations the
bulk faces satisfy. The Generalized Boundary Yang-Baxter
Equation or reflection equation is
     \newsavebox{\Bu}
     \newsavebox{\Bv}
\begin{equation}
\setlength{\unitlength}{17pt}
        \savebox{\Bu}(1.5,1){
        \begin{picture}(1.5,1)
     \put(1,1){\line(1,0){.5}}
     \put(0,0){\line(1,-1){1}}
     \put(0,0){\line(1,1){1}}
     \put(1,-1){\line(1,0){.5}}
     \multiput(1.5,-1)(0,.205){10}{\line(0,1){.15}}
     \put(1,.2){\pp{}{(r,a)}}
     \put(1,-.2){\pp{}{(u,\xi)}}
     \end{picture}}
        \savebox{\Bv}(2,1.5){
        \begin{picture}(2,1.5)
     \put(1.5,1,5){\line(1,0){.5}}
     \put(0,0){\line(1,-1){1.5}}
     \put(0,0){\line(1,1){1.5}}
     \put(1.5,-1.5){\line(1,0){.5}}
     \multiput(2,-1.5)(0,.205){15}{\line(0,1){.15}}
     \put(1.25,.2){\pp{}{(r,a)}}
     \put(1.25,-.2){\pp{}{(v,\xi)}}
     \end{picture}}
\vtop to 3\unitlength {}
\begin{picture}(3,2.5)(1,1.2)
     \put(-1,1){
\put(0,1){\line(1,1){1.5}}
\put(0,1){\line(1,-1){1}}
\put(2.5,1.5){\line(-1,1){1}}
\put(1.55,1.55){\pp{}{u-v+}}
\put(1.2,1.2){\pp{}{(q\mi p)\lambda}}
\put(.1,1){\pp{l}{\sss\uparrow}}
\put(.25,.9){\pp{b}{\sss qp}}
     }
     \put(1.4,2.5){\usebox{\Bu}}
     \put(0,0){
\put(0,1){\line(1,1){1.5}}
\put(0,1){\line(1,-1){1}}
\put(1.55,1.55){\pp{}{\mu-u-v}}
\put(1.2,1.2){\pp{}{-(p\mi 1)\lambda}}
\put(.1,1){\pp{l}{\sss\uparrow}}
\put(.25,.9){\pp{b}{\sss qp}}
     }
     \put(.9,0){\usebox{\Bv}}
\multiput(.7,3.5)(.2,0){9}{\pp{}{.}}
\put(3,3.5){\pp{l}{\;a}}
\put(3,1.5){\pp{l}{\;a}}
\put(3,-1.5){\pp{l}{\;a}}
\put(2.75,3.5){\pp{}{\alpha}}
\put(2.5,3.7){\pp{b}{b}}
\put(.5,3.7){\pp{b}{b}}
\put(-.25,2.75){\pp{}{\beta}}
\put(-1,2){\pp{r}{c\,}}
\put(-.5,1.5){\pp{}{\gamma}}
\put(0,1){\pp{r}{d\,}}
\put(.5,.5){\pp{}{\delta}}
\put(1,0){\pp{tr}{e\,}}
\put(1.75,-0.75){\pp{}{\epsilon}}
\put(2.5,-1.5){\pp{tr}{f\,}}
\put(2.75,-1.5){\pp{}{\phi}}
\end{picture}
=\;
\frac{\ss s_{1+q\mi p}^{q\,p}(u\mi v)\,s_{1-(p\mi 1)}^{q\,p}(\mu-u\mi v)
}{\ss s_{1}^{p\,q}(u\mi v)\,s_{1-(q\mi 1)}^{p\,q}(\mu-u\mi v)
}
\begin{picture}(4,2.5)(-.9,4.2)
     \put(.9,5){\usebox{\Bv}}
     \put(0,2.5){
\put(0,1.5){\line(1,1){1}}
\put(0,1.5){\line(1,-1){1.5}}
\put(1.25,1.3){\pp{}{\mu\mi{\sss (q\mi 1)}\lambda}}
\put(1.5,.95){\pp{}{-u-v}}
\put(.1,1.5){\pp{l}{\sss\uparrow}}
\put(.25,1.4){\pp{b}{\sss pq}}
     }
     \put(1.4,2.5){\usebox{\Bu}}
     \put(-1,1.5){
\put(0,1.5){\line(1,1){1}}
\put(0,1.5){\line(1,-1){1.5}}
\put(2.5,1){\line(-1,-1){1}}
\put(1.25,1.25){\pp{}{u\mi v}}
\put(.1,1.5){\pp{l}{\sss\uparrow}}
\put(.25,1.4){\pp{b}{\sss pq}}
     }
\multiput(.7,1.5)(.2,0){9}{\pp{}{.}}
\put(3,6.5){\pp{l}{\;a}}
\put(3,3.5){\pp{l}{\;a}}
\put(3,1.5){\pp{l}{\;a}}
\put(2.75,6.5){\pp{}{\alpha}}
\put(2.5,6.6){\pp{br}{b}}
\put(1.75,5.75){\pp{}{\beta}}
\put(1,5){\pp{br}{c\;}}
\put(.5,4.5){\pp{}{\gamma}}
\put(0,4){\pp{rb}{d\,}}
\put(-.5,3.5){\pp{}{\delta}}
\put(-1,3){\pp{r}{e\,}}
\put(-.25,2.25){\pp{}{\epsilon}}
\put(.5,1.3){\pp{t}{f}}
\put(2.5,1.3){\pp{t}{f\,}}
\put(2.75,1.5){\pp{}{\phi}}
\end{picture}
     \label{eq:BYBE}
\end{equation}
which is proved using the GYBE~\eqref{eq:GYBE} and the abelian
property~\eqref{eq:Abelian}.

We refer to~\cite{BPO'B96,BP01} for the boundary crossing equation.

Let's state here a property that will be of use later on. By
equation~\eqref{eq:XP}, one can fill up the triangle appearing in the
definition~\eqref{eq:B1a} of the $(1,a)$ boundary weight with any
local face operators: they will only contribute through a scalar
factor, hence,

\begin{gather}
     B^{q}_{(1,a)}\B{&&a\\
	&\gamma\\
        c\\
	&\alpha\\
		&&a}{.}={\frac{\psi_{c}^{\half}}{\psi_{a}^{\half}}} \;
		\prod_{i=1}^{q-1}\frac{1}{s_{1\mi i}^{i1}(-2u)}\,
\setlength{\unitlength}{51pt}
\vtop to 1.1\unitlength {}
\begin{picture}(1.4,1.2)(-.25,.9)
     	\thicklines
\put(0,1){\line(1,-1){1}}
\put(0,1){\line(1,1){1}}
	\thinlines
\multiput(1,0)(0,.205){10}{\line(0,1){.15}}
\multiput(0,-.2)(0,.1){25}{\pp{}{.}}
\multiput(1,-.2)(0,.1){2}{\pp{}{.}}
\multiput(1,2.2)(0,-.1){2}{\pp{}{.}}
\put(.5,2.2){\pp{}{\vec U^{q+1}_{\gamma}(c,a)^{\smash{\dag}}}}
\put(.5,-.2){\pp{}{\vec U^{q+1}_{\alpha}(c,a)}}
\put(1.1,2){\pp{l}{a}}
\put(0,1){\pp{r}{c\;\;}}
\put(1.1,0){\pp{l}{a}}
\put(.8,.2){\line(1,1){.2}}
\put(.6,.4){\line(1,1){.4}}
\put(.6,.8){\line(1,1){.4}}
\put(.2,.8){\line(1,1){.8}}
\put(.2,1.2){\line(1,-1){.8}}
\put(.6,1.2){\line(1,-1){.4}}
\put(.6,1.6){\line(1,-1){.4}}
\put(.8,1.8){\line(1,-1){.2}}
\put(.5,1.3){\p{}{\nearrow}}
\put(.5,.7){\p{}{\nwarrow}}
\put(.8,.4){\pp{}{\sss 2u\plus\lambda}}
\put(1.3,1.6){\vector(-1,0){.5}}
\put(1.3,1.6){\pp{l}{\;2u+{\sss (2q\mi 3)}\lambda}}
\put(-.3,.5){\vector(1,1){.5}}
\put(-.3,.5){\pp{t}{2u+{\sss (q\mi 1)}\lambda}}
\end{picture}\;.
		\label{eq:Spectators}
\end{gather}

\subsection{Double row transfer matrix}
\label{sec:DoubleRowTransfer}
The double row transfer matrix is given by two rows similar to the one
appearing in the torus transfer matrix, with spectral parameters $u$
for the bottom one and $\mu-u-(q-1)\lambda$ for the top one, where
$\mu$ is a fixed parameter and $q$ is the vertical fusion level.  The
boundary condition is not cyclic but fixed by the boundary weights
\eqref{eq:Bra}.

\begin{gather}
	\< \vec{a},\vec{\alpha}|\;
\vec{T}_{
	(r_{L},a_{L})|
	(r,s,\zeta)|
	(r_{R},a_{R})
	}^{pq}
	(u,\xi_{L},\xi,\xi_{R})\;
		|\vec{b},\vec{\beta}\>=\notag\\
\setlength{\unitlength}{14mm}
\vtop to 1.7\unitlength {}
\begin{picture}(9.5,1.7)(-.75,1.5)
      \multiput(0.5,0.5)(0,1){3}{\line(1,0){6.5}}
\multiput(0.5,0.5)(1,0){2}{\line(0,1){2}}
\multiput(3,0.5)(1,0){5}{\line(0,1){2}}
\multiput(0,0)(0,1){2}{
\put(0.5,.52){\pp{bl}{\sss\searrow}\pp{bl}{\sss pq}}
\multiput(2.5,0)(1,0){4}{\put(0.5,0.52){\pp{bl}{\sss\searrow}}}
\put(3,.52){\pp{bl}{\sss pq}}
\multiput(4.05,.525)(1,0){3}{\pp{bl}{\sss q}}
\put(2.25,1){\pp{}{\cdots}}
}
\put(-1,0.35){\pp{c}{a_{L}}}
\put(-.75,0.5){\pp{c}{\alpha_{L}}}
\put(0.5,0.35){\pp{c}{a_{1}}}
\put(1,0.5){\pp{c}{\alpha_{1}}}
\put(1.5,0.35){\pp{c}{a_{2}}}
\put(3,0.35){\pp{c}{a_{N}}}
\put(3.5,0.5){\pp{c}{\alpha_{N}}}
\put(4,0.35){\pp{c}{a_{N\plus1}}}
\put(4.5,0.5){\pp{c}{\alpha_{N\plus1}}}
\put(5,0.35){\pp{c}{a_{N\plus2}}}
\put(5.5,0.5){\pp{c}{\alpha_{N\plus2}}}
\put(6,0.35){\pp{c}{a_{N\plus3}}}
\put(7,0.35){\pp{c}{a_{N\plus4}}}
\put(8.25,0.5){\pp{c}{\alpha_{R}}}
\put(8.5,0.35){\pp{c}{a_{R}}}
\put(-1,2.7){\pp{c}{a_{L}}}
\put(-.75,2.5){\pp{c}{\beta_{L}}}
\put(0.5,2.7){\pp{c}{b_{1}}}
\put(1,2.5){\pp{c}{\beta_{1}}}
\put(1.5,2.7){\pp{c}{b_{2}}}
\put(3,2.7){\pp{c}{b_{N}}}
\put(3.5,2.5){\pp{c}{\beta_{N}}}
\put(4,2.7){\pp{c}{b_{N\plus1}}}
\put(4.5,2.5){\pp{c}{\beta_{N\plus1}}}
\put(5,2.7){\pp{c}{b_{N\plus2}}}
\put(5.5,2.5){\pp{c}{\beta_{N\plus2}}}
\put(6,2.7){\pp{c}{b_{N\plus3}}}
\put(7,2.7){\pp{c}{b_{N\plus4}}}
\put(8.25,2.5){\pp{c}{\beta_{R}}}
\put(8.5,2.7){\pp{c}{a_R}}
\put(1,1){\p{}{u}}
\put(3.5,1){\p{}{u}}
\put(1,2.15){\pp{r}{\bigl(\mu\mi u}}
\put(1,1.85){\pp{}{\; -\!{\sss (q\mi 1)}\lambda\bigr)}}
\put(3.5,2.15){\pp{r}{\bigl(\mu\mi u}}
\put(3.5,1.85){\pp{}{\; -\!{\sss (q\mi 1)}\lambda\bigr)}}
\put(4.5,1){\p{}{{}^{r}\! (u,\xi)}}
\put(4.75,2.15){\pp{r}{\bigl(^{\kern-.6em r\kern .5em}\mu - u}}
\put(4.5,1.85){\pp{}{-\!{\sss (q\mi 1)}\lambda,\, \xi\bigr)}}
\multiput(0,0)(0,1){2}{\put(5.5,1){\p{}{{(1,s)}}}
\put(6.5,1){\p{}{{\zeta}}}}
\put(-1,2.5){\line(1,0){.5}}
\put(-.5,2.5){\line(1,-1){1}}
\put(-.5,.5){\line(1,1){1}}
\put(-1,.5){\line(1,0){.5}}
\multiput(-1,.5)(0,.205){10}{\line(0,1){.15}}
\put(-.5,1.7){\pp{}{(r_{L},a_{L})}}
\put(-.5,1.3){\pp{}{(\mu\mi u,\xi_{L})}}
\multiput(0,0)(0,2){2}{
	\multiput(0,0)(7.5,0){2}{
\multiput(.4,.5)(-.1,0){9}{\pp{}{.}}
	}}
\put(8.5,2.5){\line(-1,0){.5}}
\put(7,1.5){\line(1,-1){1}}
\put(7,1.5){\line(1,1){1}}
\put(8.5,.5){\line(-1,0){.5}}
\multiput(8.5,.5)(0,.205){10}{\line(0,1){.15}}
\put(8,1.7){\pp{}{(r_{R},a_{R})}}
\put(8,1.3){\pp{}{(u,\xi_{R})}}
\end{picture}
         \label{eq:TpqCylinder}
\end{gather}
The GYBE~\eqref{eq:GYBE} implies that double row transfer matrices
with the same boundary conditions and boundary fields commute:
\begin{gather}
\vec{T}_{
	(r_{L},a_{L})|
	(r,s,\zeta)|
	(r_{R},a_{R})
	}^{pq}
	(u,\xi_{L},\xi,\xi_{R})\;
\vec{T}_{
	(r_{L},a_{L})|
	(r,s,\zeta)|
	(r_{R},a_{R})
	}^{pq'}
	(v,\xi_{L},\xi,\xi_{R})\;
	=\qquad\qquad
   \notag \\
\qquad\qquad
\vec{T}_{
	(r_{L},a_{L})|
	(r,s,\zeta)|
	(r_{R},a_{R})
	}^{pq'}
	(v,\xi_{L},\xi,\xi_{R})\;
\vec{T}_{
	(r_{L},a_{L})|
	(r,s,\zeta)|
	(r_{R},a_{R})
	}^{pq}
	(u,\xi_{L},\xi,\xi_{R})\;.
	 \label{eq:Commutation}
\end{gather}

\section{Fusion hierachies}
\label{sec:FusionHierarchies}\setcounter{equation}{0}
These transfer matrices fulfill a fusion hierarchy of functional
equations.  The details of these equations do depend on the type of
matrices but their structure is the same. It stems from local
properties that they all satisfy.  Let's choose a horizontal fusion
level $p$, and a fixed boundary condition among those available, namely
toroidal, cylindrical, with or without seams. Call $\vec T^{q}_{k}(u)$
the corresponding $(p,q)$-fused transfer matrix at spectral parameter
$u+k\lambda$, for $-1\leq q\leq g-2$, with $\vec T_0^{-1}$ and $\vec T_0^0$
  defined as
\begin{equation}
\vec T_0^{-1}=\vec 0\, ,
\qquad
\vec T_0^0=f_{-1}^p \vec I\quad;
\end{equation}
where
$f_q^p$ is the usual order-$N$ bulk term
\begin{equation}
f_q^p(u)=\left\{
\begin{array}{ll}
  [s_q^p(u)]^N\, , &\mbox{for the torus}\\
{}(-1)^{pN}[s_q^p(u)s_{q+p}^p(u-\mu)]^N \, ,
&\mbox{for the cylinder.}
\end{array}\right.
	\label{eq:Bulk}
\end{equation}

Then the matrices $(\vec T^{q})_{-1\leq q\leq g-3}$ fulfill a
hierarchy of functional equations
\begin{equation}
\vec T_0^{q}\:\vec T_q^{1}\:=\:
V_q\Phi_q f_q^p\:\vec T_0^{q-1}+
\tilde{V}_q f_{q-1}^p\:\vec T_0^{q+1}
\label{eq:InversionId}
\end{equation}
where $f_q^p$,  $\Phi_q$, $V_q$ and $\tilde{V}_q$, are (scalar) functions
that we are going to describe, that account for the
contributions of the bulk faces, the seams and the cylindrical
$(1,a)$-boundary conditions respectively.

%
%

The functions
$V_q$ and $\tilde{V}_q$
are trivial in the torus case, $V_q=\tilde{V}_q=1$, and on the
cylinder, they are given by
\begin{equation}
V_q=
{}\frac{s_{q-2}(2u-\mu)s_{2q+1}(2u-\mu)}
{s_{q-1}(2u-\mu)s_{2q}(2u-\mu)}\, ,
\end{equation}
and
\begin{equation}
\tilde{V}_q=
{}\frac{s_q(2u-\mu)s_{2q-1}(2u-\mu)}
{s_{q-1}(2u-\mu)s_{2q}(2u-\mu)}\, .
\end{equation}

The function $\Phi_q$ is the product of order-1 terms coming from the
seams.
As we saw in Section~\ref{sec:BoundaryWeigths}, an $(r,a)$-boundary
condition is constructed from the action of an $r$-seam on a
$(1,a)$-boundary  condition, and we count separately the type $r$
seams coming from the left and right boundaries.
If there are $K$ seams, the function  $\Phi_q$ is given by a product
of $K$ similar terms:
\begin{equation}
\Phi_q \:=\:\prod_{k=1}^{K}
\phi_q(r_k,\xi_k,u)\,.
\end{equation}
The contribution of an $(r,s,\zeta)$-seam only depends on $r$ and
$\phi_q(1,\xi,u)=1$. For $2\leq r\leq g-2$,
\begin{equation}
\phi_q(r,\xi,u)=
\left\{
\begin{array}{ll}
\phi_q^{\text{t}}(r,\xi,u)=
s_{q-r}(u+\xi)s_{q}(u+\xi)\, , &\mbox{for the torus} \\
\phi_q^{\text{t}}(r,\xi,u)\:
\phi_{q+r-1}^{\text{t}}(r,-\mu-\xi,u)
\, ,&\mbox{for the cylinder.}
\end{array}\right.
\label{eq:SeamPhi}
\end{equation}
contribution of

More generally, we have the following hierarchy of inversion identities
which can be proved by induction as in \cite{KP92}:
\begin{equation}
A_q\:\vec T_0^{q}\:\vec T_1^{q}\:=\:
B_q \:f_{-1}^p f_q^p\:\prod_{k=1}^q\Phi_k
\:\vec I +C_q\:\vec T_0^{q+1}\:\vec T_1^{q-1}
\label{eq:Hierarchy}
\end{equation}
where, in the torus case the functions $A_{q}=B_{q}=C_{q}=1$ are
trivial,
and
\begin{equation}
A_q(u) =
s_{q-1}(2u-\mu)s_{q+1}(2u-\mu)\, ,
\end{equation}
\begin{equation}
B_q(u) =
s_{-1}(2u-\mu)s_{2q+1}(2u-\mu)\, ,
\end{equation}
\begin{equation}
C_q(u) =
{}[s_{q}(2u-\mu)]^2 \,,
\end{equation}
result from
the left and right vacuum boundaries in the cylinder case.

If we further define the normalized transfer matrices
\begin{equation}
\vec t^q_0\:=\:\frac{C_q\:\vec T_1^{q-1}\:
\vec T_0^{q+1}}
{B_q f_{-1}^p f_q^p \prod_{k=1}^q\Phi_k}\, ,
\end{equation}
then the inversion identity hierarchy can be recast in the form of the
following universal \emph{thermodynamic Bethe ansatz} (TBA) functional
equation
\begin{equation}
\vec t_0^q\:\vec t_1^q\:=\:\left(\vec I +\vec t_1^{q-1}\right)
\left(\vec I +\vec t_0^{q+1}\right).
\label{eq:TBA}
\end{equation}
%
%
In deriving the TBA equation we
have used the simple properties
\begin{equation}
B_q(u) B_q(u+\lambda)\:=\:B_{q-1}(u+\lambda)B_{q+1}(u) \, ,
\end{equation}
and
\begin{equation}
\frac{C_q(u)C_q(u+\lambda)}{A_{q-1}(u+\lambda)A_{q+1}(u)}\:=\:1\,.
\end{equation}

Equations \eqref{eq:InversionId}, \eqref{eq:Hierarchy} and
\eqref{eq:TBA} give a matrix realization of the
fusion rules~\eqref{eq:F} and \eqref{eq:Ft}.

\section{Derivation}
\label{sec:Derivation}\setcounter{equation}{0}

Before we proceed to the detailed derivations of \eqref{eq:InversionId}
for the individual torus and cylinder cases, let's study the
local properties which are common to both cases.

\subsection{Local properties}
\label{sec:LocalProperties}
Firstly, let's look at how the product $\vec T_0^q\: \vec T_q^1$ is
decomposed into a sum of two terms $\vec T_0^{q-1}$ and $\vec
T_0^{q+1}$ up to scalar factors.

Because of the vertical push-through property, we can disregard the
horizontal fusion projectors and apply them later on as a wrapping of
the equation.

The product $\vec T_0^q\: \vec T_q^1$ is realized as two transfer
matrices stacked upon each other, the top one being at vertical fusion
level $1$ and the bottom one at fusion level $q$.  Consider an
arbitrary column of the torus transfer matrix ($\zeta$-type seams
excluded).  In fact, after a simple
manipulation~\eqref{eq:CylinderColumn}, the product of transfer
matrices in the cylinder case will be built up of similar columns.
There is a projector $P^{q+1}$ attached to its bottom part, realizing
the vertical fusion.  The Boltzmann weights of this column can be
written in terms of Temperley-Lieb operators,
%
\begin{equation}
X_{j\plus q\mi 1}(v)\ldots X_{j\plus 1}\left(v+(q-2)\lambda\right)
X_{j}\left(v+(q-1)\lambda\right)\:P_{j+1}^{q+1}\:X_{j\mi 1}(v+q\lambda)
	\label{eq:Column}
\end{equation}
  with $j$ an arbitrary label and $v$ the spectral parameter involved
  in that particular column, for example $v=u-k\lambda$ for a typical
  bulk face and $v=u+\xi-k\lambda$ for a face in an $r$-seam.  Because
  an $s$-type seam is the braid limit of an $r$-type seam, we don't
  lose any generality in considering only $r$-type seams.  It is easy
  to see that the following arguments can be applied also to
  $\zeta$-type seams and that their contribution is trivial.

  We duplicate the projector and insert between its two copies the
  following identity
\begin{equation}
\frac{1}{S_{q+1}}\left(S_q X_j(\lambda)+X_j(-q\lambda)\right)\:=\: \vec{I}\
\end{equation}
\begin{equation}
\setlength{\unitlength}{9mm}
\begin{picture}(8,5.5)
\put(-3.75,3){$    \eqref{eq:Column}
\displaystyle =$}
\put(-1.25,0.3){
\begin{picture}(4,5.5)
\multiput(0.5,0.5)(1,0){2}{\line(0,1){5}}
\multiput(0.5,0.5)(0,1){6}{\line(1,0){1}}
\multiput(0.48,0.58)(3,0){1}{\pp{bl}{\searrow}}
\multiput(0.48,4.58)(3,0){1}{\pp{bl}{\searrow}}
\multiput(0.48,3.58)(3,0){1}{\pp{bl}{\searrow}}
\put(1,1){\pp{c}{v}}
\put(0.6,5){\pp{l}{v+q\lambda}}
\put(0.6,4.2){\pp{l}{v+}}\put(1.5,3.9){\pp{r}{(q-1)\lambda}}
%
%
\multiput(1.5,1.5)(1,0){1}{\pp{}{\bullet}}
\multiput(1.5,3.5)(1,0){1}{\pp{}{\bullet}}
\multiput(1.5,2.5)(1,0){1}{\pp{}{\bullet}}
%
%
\multiput(1.5,1.5)(1,0){2}{\line(0,1){2}}
\put(1.5,1.5){\line(1,-2){0.5}} \put(2,0.5){\line(1,2){0.5}}
\put(1.5,3.5){\line(1,2){0.5}} \put(2,4.5){\line(1,-2){0.5}}
\multiput(1.5,0.5)(0.1,0){5}{\pp{}{.}}
\multiput(1.5,4.5)(0.1,0){5}{\pp{}{.}}
\put(2.05,2.5){\pp{c}{P^{q+1}_{j+1}}}
%
%
\multiput(0.7,0.7)(0.1,0.1){2}{\pp{}{.}}
\multiput(1.5,1.5)(-0.1,-0.1){4}{\pp{}{.}}
\multiput(0.5,0.5)(-0.1,-0.1){4}{\pp{}{.}}
\put(0,0){\pp{c}{j+q-1}}
\multiput(0.5,1.5)(0.1,0.1){10}{\pp{}{.}}
\multiput(0.5,1.5)(-0.1,-0.1){4}{\pp{}{.}}
\put(0,1){\pp{c}{j+q-2}}
\multiput(0.5,3.5)(-0.1,-0.1){4}{\pp{}{.}}
\put(0,3){\pp{c}{j}}
\multiput(0.5,4.5)(-0.1,-0.1){4}{\pp{}{.}}
\put(0,4){\pp{c}{j-1}}
\end{picture}}
\put(1.7,3){$\displaystyle =$}
  \put(3.2,0.3){
\begin{picture}(5,5.5)
\multiput(0.5,0.5)(1,0){2}{\line(0,1){5}}
\multiput(0.5,0.5)(0,1){6}{\line(1,0){1}}
\multiput(0.48,0.58)(3,0){1}{\pp{bl}{\searrow}}
\multiput(0.48,4.58)(3,0){1}{\pp{bl}{\searrow}}
\multiput(0.48,3.58)(3,0){1}{\pp{bl}{\searrow}}
\put(1,1){\pp{c}{v}}
\put(0.6,5){\pp{l}{v+q\lambda}}
\put(0.6,4.2){\pp{l}{v+}}\put(1.5,3.9){\pp{r}{(q-1)\lambda}}
%
%
\multiput(1.5,1.5)(1,0){1}{\pp{}{\bullet}}
\multiput(1.5,3.5)(1,0){1}{\pp{}{\bullet}}
\multiput(1.5,2.5)(1,0){1}{\pp{}{\bullet}}
%
%
\multiput(1.5,1.5)(1,0){2}{\line(0,1){2}}
\put(1.5,1.5){\line(1,-2){0.5}}
\put(2,0.5){\line(1,2){0.5}}
\put(1.5,3.5){\line(1,2){0.5}}
\put(2,4.5){\line(1,-2){0.5}}
\multiput(1.5,0.5)(0.1,0){5}{\pp{}{.}}
\multiput(1.5,4.5)(0.1,0){5}{\pp{}{.}}
\put(2.05,2.5){\pp{c}{P^{q+1}_{j+1}}}
%
%
\multiput(6.5,1.5)(1,0){2}{\line(0,1){2}}
\put(6.5,1.5){\line(1,-2){0.5}}
\put(7,0.5){\line(1,2){0.5}}
\put(6.5,3.5){\line(1,2){0.5}}
\put(7,4.5){\line(1,-2){0.5}}
\multiput(2,0.5)(0.1,0){50}{\pp{}{.}}
\multiput(6.5,4.5)(0.1,0){5}{\pp{}{.}}
\put(7.05,2.5){\pp{c}{P^{q+1}_{j+1}}}
\multiput(6.5,1.5)(0,2){2}{\pp{}{\bullet}}
%
%
\put(-0.3,3){\pp{c}{\displaystyle\frac{1}{S_{q+1}}\times}}
\multiput(1.5,5.5)(0.1,0){55}{\pp{}{.}}
\multiput(2.5,1.5)(0.1,0){40}{\pp{}{.}}
\multiput(2.5,3.5)(0.1,0){40}{\pp{}{.}}
\put(2.5,1.5){\pp{}{\bullet}}
\multiput(2,4.5)(0.1,0){3}{\pp{}{.}}
\multiput(2.8,4.5)(0.1,0){3}{\pp{}{.}}
\multiput(4.1,4.5)(0.1,0){3}{\pp{}{.}}
\multiput(4.7,4.5)(0.1,0){3}{\pp{}{.}}
\multiput(6,4.5)(0.1,0){3}{\pp{}{.}}
\put(2.6,4.5){\pp{c}{\Bigg(\displaystyle S_q}}
\put(3,4.5){\line(1,2){0.5}}
\put(4,4.5){\line(-1,2){0.5}}
\put(3,4.5){\line(1,-2){0.5}}
\put(4,4.5){\line(-1,-2){0.5}}
\put(2.5,3.5){\pp{}{\bullet}}
\put(3.5,3.5){\pp{}{\bullet}}
\put(3.5,4.5){\pp{c}{\lambda}}
\put(3.1,4.4){\pp{bl}{\downarrow}}
\put(4.5,4.5){\pp{c}{+}}
\put(5,4.5){\line(1,2){0.5}}
\put(6,4.5){\line(-1,2){0.5}}
\put(5,4.5){\line(1,-2){0.5}}
\put(6,4.5){\line(-1,-2){0.5}}
\put(5.5,3.5){\pp{}{\bullet}}
\put(5.5,4.5){\pp{c}{-q\lambda}}
\put(5.1,4.4){\pp{bl}{\downarrow}}
\put(6.3,4.5){\pp{c}{\Bigg)}}
\end{picture}
  }
\end{picture}
\label{YqX}
\end{equation}

\begin{equation}
\setlength{\unitlength}{9mm}
\vtop to 2.5 \unitlength {}
\begin{picture}(4,2)(0,3)
\put(-3,3){$\displaystyle =$}
\put(-0.8,0.3){
\begin{picture}(5,5.5)
\multiput(0.5,0.5)(1,0){2}{\line(0,1){5}}
\multiput(0.5,0.5)(0,1){6}{\line(1,0){1}}
\multiput(0.48,0.58)(3,0){1}{\pp{bl}{\searrow}}
\multiput(0.48,4.58)(3,0){1}{\pp{bl}{\searrow}}
\multiput(0.48,3.58)(3,0){1}{\pp{bl}{\searrow}}
\put(1,1){\pp{c}{v}}
\put(0.6,5){\pp{l}{v+q\lambda}}
\put(0.6,4.2){\pp{l}{v+}}\put(1.5,3.9){\pp{r}{(q-1)\lambda}}
%
%
\multiput(1.5,1.5)(1,0){1}{\pp{}{\bullet}}
\multiput(1.5,3.5)(1,0){1}{\pp{}{\bullet}}
\multiput(1.5,2.5)(1,0){1}{\pp{}{\bullet}}
\multiput(1.5,4.5)(0.5,0){1}{\pp{}{\bullet}}
\put(0.35,0){
\begin{picture}(4,5.5)
%
\multiput(2,1.5)(1,0){2}{\line(0,1){2}}
\put(2,1.5){\line(1,-2){0.5}}
\put(2.5,0.5){\line(1,2){0.5}}
\put(2,3.5){\line(1,2){0.5}}
\put(2.5,4.5){\line(1,-2){0.5}}
\multiput(1.5,0.5)(0.1,0){10}{\pp{}{.}}
\put(2.55,2.5){\pp{c}{P^{q+1}_{j+1}}}
\end{picture} }
%
%
\multiput(1.5,1.5)(1,0){2}{\line(0,1){2}}
\put(1.5,1.5){\line(1,-2){0.5}}
\put(2,0.5){\line(1,2){0.5}}
\put(1.5,3.5){\line(1,2){0.5}}
\put(2,4.5){\line(1,-2){0.5}}
\multiput(1.5,0.5)(0.1,0){5}{\pp{}{.}}
\multiput(1.5,4.5)(0.1,0){5}{\pp{}{.}}
\put(2.05,2.5){\pp{c}{P^{q+1}_{j+1}}}
%
%
%
\put(-0.3,3){\pp{c}{\displaystyle\frac{S_q}{S_{q+1}}\times}}
\multiput(1.5,5.5)(0.1,0){10}{\pp{}{.}}
\multiput(1.5,1.5)(0.1,0){0}{\pp{}{.}}
\multiput(1.5,2.5)(0.1,0){0}{\pp{}{.}}
\multiput(1.5,3.5)(0.1,0){0}{\pp{}{.}}
\put(0.35,0){
\begin{picture}(4,5.5)
\put(1.5,4.5){\line(1,2){0.5}}
\put(2.5,4.5){\line(-1,2){0.5}}
\put(1.5,4.5){\line(1,-2){0.5}}
\put(2.5,4.5){\line(-1,-2){0.5}}
\put(2,4.5){\pp{c}{\lambda}}
\put(1.6,4.4){\pp{bl}{\downarrow}}
\end{picture} }
\end{picture}}
\put(3.5,3){$\displaystyle +$}
  \put(4,0.3){
\begin{picture}(5,5.5)
\multiput(0.5,0.5)(1,0){2}{\line(0,1){5}}
\multiput(0.5,0.5)(0,1){6}{\line(1,0){1}}
\multiput(0.48,0.58)(3,0){1}{\pp{bl}{\searrow}}
\multiput(0.48,4.58)(3,0){1}{\pp{bl}{\searrow}}
\multiput(0.48,3.58)(3,0){1}{\pp{bl}{\searrow}}
\put(1,1){\pp{c}{v}}
\put(0.6,5){\pp{l}{v+q\lambda}}
\put(0.6,4.2){\pp{l}{v+}}\put(1.5,3.9){\pp{r}{(q-1)\lambda}}
%
%
\multiput(1.5,1.5)(1,0){1}{\pp{}{\bullet}}
\multiput(1.5,3.5)(1,0){1}{\pp{}{\bullet}}
\multiput(1.5,2.5)(1,0){1}{\pp{}{\bullet}}
\multiput(1.5,4.5)(0.5,0){1}{\pp{}{\bullet}}
%
%
\multiput(1.5,1.5)(1,0){2}{\line(0,1){3}}
\put(1.5,1.5){\line(1,-2){0.5}}
\put(2,0.5){\line(1,2){0.5}}
\put(1.5,4.5){\line(1,2){0.5}}
\put(2,5.5){\line(1,-2){0.5}}
\multiput(1.5,0.5)(0.1,0){5}{\pp{}{.}}
\multiput(1.5,5.5)(0.1,0){5}{\pp{}{.}}
\put(2.05,3.25){\pp{c}{P^{q+2}_{j}}}
\end{picture}}
\end{picture}
\label{YqX2}
\end{equation}

%
%
The projector $P_j^{q+2}$ in the second term of \eqref{YqX2} is
obtained by the definition~\eqref{eq:DefP}.  Thus, this term gives us
the column which appears in $\vec T_0^{q+1}$ in the functional
equation \eqref{eq:InversionId}.  By pushing the projector through
horizontally in the product of transfer matrices, we can make it
appear in between all columns and because of the cyclic boundary
condition (and a similar argument in the cylinder case), we finally
obtain a term which is proportional to $\vec T_0^{q+1}$.

We are now going to prove that the first term of \eqref{YqX2} yields a
term proportional to $\vec T_0^{q-1}$.

The product $\vec T_0^q\: \vec T_q^1$ involves a whole row of columns
such as the LHS of~\eqref{YqX}, hence the columns of Boltzmann weights
occur with a fusion projector $P^{q+1}$ between each of them.  We
can use the push-through property~\eqref{eq:PushThru} to remove
all but one of these projectors one by one:
\begin{equation}
\setlength{\unitlength}{9mm}
\vtop to 2\unitlength {}
     \begin{picture}(3,3)(-.5,2.5)
\multiput(0.5,0.5)(1,0){2}{\line(0,1){5}}
\multiput(0.5,0.5)(0,1){2}{\line(1,0){1}}
\multiput(0.5,3.5)(0,1){3}{\line(1,0){1}}
\multiput(0.48,0.58)(3,0){1}{\pp{bl}{\searrow}}
\multiput(0.48,4.58)(3,0){1}{\pp{bl}{\searrow}}
\multiput(0.48,3.58)(3,0){1}{\pp{bl}{\searrow}}
\put(0.6,5){\pp{l}{v+q\lambda}}
\put(0.6,4.2){\pp{l}{v+}}\put(1.5,3.9){\pp{r}{{\sss (q\mi 1)}\lambda}}
\put(1,2.5){\p{c}{\uparrow}}
\put(1,1){\pp{c}{v}}
\multiput(1.5,1.5)(1,0){1}{\pp{}{\bullet}}
\multiput(1.5,3.5)(1,0){1}{\pp{}{\bullet}}
\multiput(.5,1.5)(1,0){1}{\pp{}{\bullet}}
\multiput(.5,3.5)(1,0){1}{\pp{}{\bullet}}
%
\put(-2,0){
\multiput(1.5,1.5)(1,0){2}{\line(0,1){2}}
\put(1.5,1.5){\line(1,-2){0.5}} \put(2,0.5){\line(1,2){0.5}}
\put(1.5,3.5){\line(1,2){0.5}} \put(2,4.5){\line(1,-2){0.5}}
\multiput(2,0.5)(0.1,0){5}{\pp{}{.}}
\multiput(2,4.5)(0.1,0){5}{\pp{}{.}}
\put(2.05,2.5){\pp{c}{P^{q+1}}}
}
\put(0,0){
\multiput(1.5,1.5)(1,0){2}{\line(0,1){2}}
\put(1.5,1.5){\line(1,-2){0.5}} \put(2,0.5){\line(1,2){0.5}}
\put(1.5,3.5){\line(1,2){0.5}} \put(2,4.5){\line(1,-2){0.5}}
\multiput(1.5,0.5)(0.1,0){5}{\pp{}{.}}
\multiput(1.5,4.5)(0.1,0){5}{\pp{}{.}}
\put(2.05,2.5){\pp{c}{P^{q+1}}}
}
\end{picture}
\; = \;
     \begin{picture}(3,3)(.5,2.5)
\multiput(0.5,0.5)(1,0){2}{\line(0,1){5}}
\multiput(0.5,0.5)(0,1){2}{\line(1,0){1}}
\multiput(0.5,3.5)(0,1){3}{\line(1,0){1}}
\multiput(0.48,0.58)(3,0){1}{\pp{bl}{\searrow}}
\multiput(0.48,4.58)(3,0){1}{\pp{bl}{\searrow}}
\multiput(0.48,3.58)(3,0){1}{\pp{bl}{\searrow}}
\put(0.6,5){\pp{l}{v+q\lambda}}
\put(1,4){\pp{c}{v}}
\put(1,2.5){\p{c}{\downarrow}}
\put(0.6,1.2){\pp{l}{v+}}\put(1.5,.9){\pp{r}{{\sss (q\mi 1)}\lambda}}
\multiput(1.5,1.5)(1,0){1}{\pp{}{\bullet}}
\multiput(1.5,3.5)(1,0){1}{\pp{}{\bullet}}
%
\put(0,0){
\multiput(1.5,1.5)(1,0){2}{\line(0,1){2}}
\put(1.5,1.5){\line(1,-2){0.5}} \put(2,0.5){\line(1,2){0.5}}
\put(1.5,3.5){\line(1,2){0.5}} \put(2,4.5){\line(1,-2){0.5}}
\multiput(1.5,0.5)(0.1,0){15}{\pp{}{.}}
\multiput(1.5,4.5)(0.1,0){15}{\pp{}{.}}
\put(2.05,2.5){\pp{c}{P^{q+1}}}
}
\put(1,0){
\multiput(1.5,1.5)(1,0){2}{\line(0,1){2}}
\put(1.5,1.5){\line(1,-2){0.5}} \put(2,0.5){\line(1,2){0.5}}
\put(1.5,3.5){\line(1,2){0.5}} \put(2,4.5){\line(1,-2){0.5}}
\put(2.05,2.5){\pp{c}{P^{q+1}}}
}
\end{picture}
\; = \;
     \begin{picture}(2,3)(-.5,2.5)
\multiput(0.5,0.5)(1,0){2}{\line(0,1){5}}
\multiput(0.5,0.5)(0,1){2}{\line(1,0){1}}
\multiput(0.5,3.5)(0,1){3}{\line(1,0){1}}
\multiput(0.48,0.58)(3,0){1}{\pp{bl}{\searrow}}
\multiput(0.48,4.58)(3,0){1}{\pp{bl}{\searrow}}
\multiput(0.48,3.58)(3,0){1}{\pp{bl}{\searrow}}
\put(0.6,5){\pp{l}{v+q\lambda}}
\put(0.6,4.2){\pp{l}{v+}}\put(1.5,3.9){\pp{r}{{\sss (q\mi 1)}\lambda}}
\put(1,2.5){\p{c}{\uparrow}}
\put(1,1){\pp{c}{v}}
\multiput(.5,1.5)(1,0){1}{\pp{}{\bullet}}
\multiput(.5,3.5)(1,0){1}{\pp{}{\bullet}}
%
\put(-2,0){
\multiput(1.5,1.5)(1,0){2}{\line(0,1){2}}
\put(1.5,1.5){\line(1,-2){0.5}} \put(2,0.5){\line(1,2){0.5}}
\put(1.5,3.5){\line(1,2){0.5}} \put(2,4.5){\line(1,-2){0.5}}
\multiput(2,0.5)(0.1,0){5}{\pp{}{.}}
\multiput(2,4.5)(0.1,0){5}{\pp{}{.}}
\put(2.05,2.5){\pp{c}{P^{q+1}}}
}
\end{picture}
     \label{eq:PushThru2}
\end{equation}

So for cyclic boundary conditions, the projector on the left of the
$+\lambda$ face in the first term of \eqref{YqX2} can be discarded and
the one on its left will be the only remaining projector in the row.  We
will see that the same argument is also valid for cylindrical
boundary conditions.

  We now make use
of the \emph{contracting} property of the local face projector
$X_{j}(\lambda)$:
\begin{equation}
\setlength{\unitlength}{4mm}
\vtop to 1.9 \unitlength {}
\begin{picture}(8,2.6)(0,1.5)
\put(-3.5,0){
\begin{picture}(3,4)
\put(2,0){\line(-1,1){1}}
\put(2,0){\line(1,1){1}}
\put(2,2){\line(1,-1){1}}
\put(2,2){\line(-1,-1){1}}
\put(2,2){\line(-1,1){1}}
\put(1,3){\line(-1,-1){1}}
\put(1,1){\line(-1,1){1}}
\multiput(1,3)(1,-1){2}{\line(1,1){1}}
\put(2,4){\line(1,-1){1}}
\put(2,3){\pp{}{\lambda}}
\put(1,2){\pp{}{u+\lambda}}
\put(2,1){\pp{}{u}}
\multiput(0,-0.2)(0,0.2){23}{\pp{}{.}}
\multiput(1,-0.2)(0,0.2){6}{\pp{}{.}}
\multiput(2,-0.2)(0,0.2){1}{\pp{}{.}}
\multiput(3,-0.2)(0,0.2){23}{\pp{}{.}}
\multiput(1,4.2)(0,-0.2){6}{\pp{}{.}}
\multiput(2,4.2)(0,-0.2){2}{\pp{}{.}}
\multiput(1.76,0.25)(0,2){2}{\pp{bl}{\sss \to}}
\put(0.76,1.25){\pp{bl}{\sss \to}}
\put(1,-0.5){\pp{}{j-1}}
\put(2,-0.5){\pp{}{j}}
\end{picture} }
\put(1,1.5){$\displaystyle =$}
\put(2.5,1.5){$\displaystyle
s_1(u)s_1(-u)
$}
\put(8,0){
\begin{picture}(3,4)
\put(2,2){\line(-1,-1){1}}
\put(2,2){\line(-1,1){1}}
\put(1,3){\line(-1,-1){1}}
\put(1,1){\line(-1,1){1}}
\multiput(1,3)(1,-1){2}{\line(1,1){1}}
\put(2,4){\line(1,-1){1}}
\put(2,3){\pp{}{\lambda}}
\put(1,2){\pp{}{\lambda}}
\multiput(0,-0.2)(0,0.2){23}{\pp{}{.}}
\multiput(1,-0.2)(0,0.2){6}{\pp{}{.}}
\multiput(2,-0.2)(0,0.2){11}{\pp{}{.}}
\multiput(3,-0.2)(0,0.2){23}{\pp{}{.}}
\multiput(1,4.2)(0,-0.2){6}{\pp{}{.}}
\multiput(2,4.2)(0,-0.2){2}{\pp{}{.}}
\multiput(1.76,2.25)(0,2){1}{\pp{bl}{\sss \to}}
\put(0.76,1.25){\pp{bl}{\sss \to}}
\put(1,-0.5){\pp{}{j-1}}
\put(2,-0.5){\pp{}{j}}
\end{picture} }
\end{picture}
\label{collapse1}
\end{equation}
so that the two faces in the top rows of the first term of
\eqref{YqX2} collapse into a scalar under the propagation of the
contractor:
\begin{equation}
\setlength{\unitlength}{9mm}
\begin{picture}(8,6)
\put(-1.65,0.3){
\begin{picture}(4,5.5)
\multiput(3,0.5)(0,1){2}{\line(1,0){0.2}}
\multiput(3,3.5)(0,1){3}{\line(1,0){0.2}}
\multiput(3,0.5)(0,1){1}{\line(0,1){5}}
\multiput(0.5,1.5)(0,1){1}{\pp{}{\bullet}}
\multiput(0.5,3.5)(0,1){2}{\pp{}{\bullet}}
\multiput(3,1.5)(0,1){1}{\pp{}{\bullet}}
\multiput(3,3.5)(0,1){2}{\pp{}{\bullet}}
\multiput(3,0.5)(-0.1,0){5}{\pp{}{.}}
\multiput(3,4.5)(-0.1,0){5}{\pp{}{.}}
\multiput(3,5.5)(-0.1,0){10}{\pp{}{.}}
\multiput(0.5,0.5)(1,0){2}{\line(0,1){5}}
\multiput(0.2,0.5)(0,1){2}{\line(1,0){1.3}}
\multiput(0.2,3.5)(0,1){3}{\line(1,0){1.3}}
\multiput(0.48,0.58)(3,0){1}{\pp{bl}{\searrow}}
\multiput(0.48,4.58)(3,0){1}{\pp{bl}{\searrow}}
\multiput(0.48,3.58)(3,0){1}{\pp{bl}{\searrow}}
\put(1,2.5){\p{}{\uparrow}}
\put(1,1){\pp{c}{v}}
\put(0.6,5){\pp{l}{v+q\lambda}}
\put(0.6,4.2){\pp{l}{v+}}\put(1.5,3.9){\pp{r}{(q-1)\lambda}}
\put(0.63,5.52){\pp{br}{b_{i\mi 1}}}
\put(0.45,3.5){\pp{br}{c_{i\mi 1}}}
\put(0.63,.4){\pp{tr}{a_{i\mi 1}}}
\put(3,5.55){\pp{b}{b_i}}
\put(1.5,5.6){\pp{b}{b_i}}
\put(2.8,4.5){\pp{b}{d_i}}
\put(1.7,3.6){\pp{b}{c_{i}}}
\put(3,.4){\pp{t}{a_i}}
\put(1.5,.4){\pp{t}{a_i}}
\multiput(1.5,1.5)(1,0){1}{\pp{}{\bullet}}
\multiput(1.5,3.5)(1,0){1}{\pp{}{\bullet}}
\multiput(1.5,4.5)(0.5,0){1}{\pp{}{\bullet}}
%
%
\multiput(2,1.5)(1,0){2}{\line(0,1){2}}
\put(2,1.5){\line(1,-2){0.5}}
\put(2.5,0.5){\line(1,2){0.5}}
\put(2,3.5){\line(1,2){0.5}}
\put(2.5,4.5){\line(1,-2){0.5}}
\multiput(1.5,0.5)(0.1,0){10}{\pp{}{.}}
\multiput(2.5,4.5)(1,0){1}{\pp{}{\bullet}}
\multiput(2,1.5)(0,1){1}{\pp{}{\bullet}}
\put(2.55,2.5){\pp{c}{P^{q+1}_{j+1}}}
%
%
%
\put(-0.75,3){\pp{c}{\displaystyle\frac{S_q}{S_{q+1}}}}
\multiput(1.5,5.5)(0.1,0){5}{\pp{}{.}}
\multiput(1.5,1.5)(0.1,0){5}{\pp{}{.}}
\multiput(1.5,3.5)(0.1,0){5}{\pp{}{.}}
\put(1.5,4.5){\line(1,2){0.5}}
\put(2.5,4.5){\line(-1,2){0.5}}
\put(1.5,4.5){\line(1,-2){0.5}}
\put(2.5,4.5){\line(-1,-2){0.5}}
\put(2,3.5){\pp{}{\bullet}}
\put(2,4.5){\pp{c}{\lambda}}
\put(1.6,4.4){\pp{bl}{\downarrow}}
\end{picture}}
\put(1.8,3){$\displaystyle=$}
\put(2.3,3){$\displaystyle
-s_{q}(v)s_{q-2}(v)
\frac{S_q}{S_{q+1}}
$}
\put(7.2,0.3){
\begin{picture}(4,5.5)
\put(-1.2,0){
\begin{picture}(4,5.5)
\multiput(.5,.5)(0,1){6}{\line(-1,0){0.3}}
\put(0.5,0.5){\line(0,1){5}}
\multiput(0.5,1.5)(0,1){4}{\pp{}{\bullet}}
\put(0.25,3.55){\pp{b}{c_{i\!-\!1}}}
\put(2,2){\pp{}{\uparrow}}
\end{picture} }
\multiput(0,0)(0,1){4}{
\multiput(0.5,0.5)(-0.1,0){10}{\pp{}{.}} }
\multiput(0.5,4.5)(0,0.1){11}{\pp{}{.}}
\put(-2.2,0){
\begin{picture}(2,2)
\multiput(1.5,5.5)(0.1,0){5}{\pp{}{.}}
\put(1.65,5.55){\pp{b}{b_{i\!-\!1}}}
\put(1.5,4.5){\line(1,2){0.5}}
\put(2.5,4.5){\line(-1,2){0.5}}
\put(1.5,4.5){\line(1,-2){0.5}}
\put(2.5,4.5){\line(-1,-2){0.5}}
\put(2,3.5){\pp{}{\bullet}}
\put(2,4.5){\pp{c}{\lambda}}
\put(1.6,4.4){\pp{bl}{\downarrow}}
\end{picture} }
\multiput(2.5,0.5)(0,1){6}{\line(1,0){0.2}}
\multiput(2.5,0.5)(0,1){1}{\line(0,1){5}}
\multiput(2.5,1.5)(0,1){4}{\pp{}{\bullet}}
\multiput(0.5,1.5)(0,1){3}{\pp{}{\bullet}}
\multiput(2.5,0.5)(-0.1,0){5}{\pp{}{.}}
\multiput(2.5,4.5)(-0.1,0){5}{\pp{}{.}}
\multiput(2.5,5.5)(-0.1,0){20}{\pp{}{.}}
\put(0.65,4.55){\pp{b}{b_i}}
\put(2.6,5.55){\pp{b}{b_i}}
\put(1.5,4.57){\pp{b}{d_i}}
\put(2.38,4.57){\pp{b}{d_i}}
\put(1.8,3.5){\pp{b}{c_{i}}}
\put(-.5,.4){\pp{t}{a_{i\mi 1}}}
\put(.5,.4){\pp{t}{a_{i\mi 1}}}
\put(1.5,.4){\pp{t}{a_i}}
\put(2.5,.4){\pp{t}{a_i}}
\multiput(0.5,0.5)(1,0){1}{\line(0,1){4}}
\multiput(1.5,0.5)(1,0){1}{\line(0,1){4}}
\multiput(0.5,0.5)(0,1){5}{\line(1,0){1}}
\multiput(0.48,0.58)(3,0){1}{\pp{bl}{\searrow}}
\multiput(0.48,3.58)(0,-1){2}{\pp{bl}{\searrow}}
\put(1,1){\pp{c}{v}}
\put(1,4){\pp{c}{\lambda}}
\put(0.6,3.2){\pp{l}{v+}}\put(1.5,2.9){\pp{r}{(q-2)\lambda}}
\multiput(1.5,1.5)(1,0){1}{\pp{}{\bullet}}
\multiput(1.5,3.5)(1,0){1}{\pp{}{\bullet}}
\multiput(1.5,2.5)(1,0){1}{\pp{}{\bullet}}
\multiput(1.5,4.5)(0.5,0){2}{\pp{}{\bullet}}
%
%
\multiput(1.5,1.5)(1,0){2}{\line(0,1){2}}
\put(1.5,1.5){\line(1,-2){0.5}} \put(2,0.5){\line(1,2){0.5}}
\put(1.5,3.5){\line(1,2){0.5}} \put(2,4.5){\line(1,-2){0.5}}
\multiput(1.5,0.5)(0.1,0){5}{\pp{}{.}}
\multiput(1.5,4.5)(0.1,0){5}{\pp{}{.}}
\put(2.05,2.5){\pp{c}{P^{q+1}_{j+1}}}
\end{picture} }
\end{picture}
\label{YqX3}
\vspace{-10pt}
\end{equation}
and the newly appeared contractor further collapses the top two faces
of the next column on the left.  Reapplying the procedure to the rest
of the columns on the left and using the cyclic boundary conditions,
we finally collapse all of the top two rows.  What is left is a
scalar contribution $-s_{q}(u)s_{q-2}(u)$ for each column at spectral
parameter $u$, a row of faces with spectral parameter $\lambda$ and
the local face operator $X_{j}(\lambda)$ at the right of the projector
$P^{q+1}$.  But the Boltzmann weight of a face at spectral parameter
$\lambda$ is simply
\begin{equation}
     W\W{d&c\\
     a&b}{|\lambda}=\frac{\psi_{a}^{\half}\psi_{c}^{\half}}
     {\psi_{b}^{\half}\psi_{d}^{\half}}\;\delta_{bd},
     \label{eq:lambda}
\end{equation}
hence the top row disappears, leaving
\begin{equation}
\setlength{\unitlength}{9mm}
\vtop to 2\unitlength {}
\begin{picture}(3.5,3.5)(0,1.5)
     \multiput(0,0)(2.5,0){2}{
\multiput(0,0)(1,0){2}{\line(0,1){4}}
\multiput(0,0)(0,1){5}{\line(1,0){1}}
\multiput(0,0)(0,1){4}{\put(.05,.05){\pp{bl}{\sss \searrow}}}
\multiput(0,1)(1,0){2}{\multiput(0,0)(0,1){3}{\pp{}{\bullet}}}
\put(.5,3.5){\pp{}{\lambda}}
\put(.5,2.4){\pp{}{{\sss (q\mi 2)}\lambda}}
\put(.5,1.5){\pp{}{\uparrow}}
}
\put(.5,2.6){\pp{r}{v+}}
\put(.5,.5){\pp{}{v}}
\put(2.5,0)	{
\put(.5,2.6){\pp{r}{w+}}
\put(.5,.5){\pp{}{w}}
		}
\multiput(0,0)(0,1){5}{\line(-1,0){.25}}
\multiput(3.5,0)(0,1){5}{\line(1,0){.25}}
\put(1,4){\pp{}{\bullet}}
\put(1.5,0)	{
\put(0,4){\line(-1,-2){.5}}
\put(0,4){\line(1,-2){.5}}
\put(0,0){\line(-1,2){.5}}
\put(0,0){\line(1,2){.5}}
\put(.5,1){\line(0,1){2}}
\put(0,2){\pp{}{P^{q\plus 1}_{j\plus 1}}}
\put(0,4){\pp{}{\bullet}}
		}
\multiput(2,1)(0,1){3}{\pp{}{\bullet}}
\multiput(2,1)(0,1){3}	{\multiput(.1,0)(.1,0){4}{\pp{}{.}}
			}
\multiput(1.1,4)(.1,0){4}{\pp{}{.}}
\put(1.5,4)
{
\put(0,0){\line(1,2){.5}}
\put(0,0){\line(1,-2){.5}}
\put(1,0){\line(-1,2){.5}}
\put(1,0){\line(-1,-2){.5}}
\put(.5,0){\pp{}{\lambda}}
\put(.1,-.1){\pp{bl}{\downarrow}}
}
\put(0,4.2){\pp{b}{b_{i}}}
\put(1,4.2){\pp{br}{d_{i}}}
\put(2,5.2){\pp{b}{b_{i}}}
\put(2.5,4.2){\pp{bl}{b_{i\plus 1}}}
\put(3.5,4.2){\pp{bl}{b_{i\plus 2}}}
\put(0,-.2){\pp{t}{a_{i\mi 1}}}
\put(1,-.2){\pp{t}{a_{i}}}
\multiput(1.1,0)(.1,0){14}{\pp{}{.}}
\put(2.5,-.2){\pp{t}{a_{i}}}
\put(3.5,-.2){\pp{t}{a_{i\plus 1}}}
\end{picture}
\;\; = \; \frac{\psi_{d_{i}}}{\psi_{b_{i}}}\;\;
\begin{picture}(3,4)(0,1.5)
     \multiput(0,0)(2,0){2}{
\multiput(0,0)(1,0){2}{\line(0,1){3}}
\multiput(0,0)(0,1){4}{\line(1,0){1}}
\multiput(0,0)(0,1){3}{\put(.05,.05){\pp{bl}{\sss \searrow}}}
\multiput(0,1)(1,0){2}{\multiput(0,0)(0,1){2}{\pp{}{\bullet}}}
\put(.5,2.4){\pp{}{{\sss (q\mi 2)}\lambda}}
\put(.5,1.5){\pp{}{\uparrow}}
}
\put(.5,2.6){\pp{r}{v+}}
\put(.5,.5){\pp{}{v}}
\put(2,0)	{
\put(.5,2.6){\pp{r}{w+}}
\put(.5,.5){\pp{}{w}}
		}
\multiput(0,0)(0,1){4}{\line(-1,0){.25}}
\multiput(3,0)(0,1){4}{\line(1,0){.25}}
\put(1.5,0)	{
\put(0,4){\line(-1,-2){.5}}
\put(0,4){\line(1,-2){.5}}
\put(0,0){\line(-1,2){.5}}
\put(0,0){\line(1,2){.5}}
\put(.5,1){\line(0,1){2}}
\put(0,2){\pp{}{P^{q\plus 1}_{j\plus 1}}}
\put(0,4){\pp{}{\bullet}}
		}
\put(0,3.2){\pp{b}{b_{i\mi 1}}}
\put(1.5,4.2){\pp{br}{d_{i}}}
\put(1,3.2){\pp{br}{b_{i}}}
\put(2,3.2){\pp{bl}{b_{i}}}
\put(3,3.2){\pp{bl}{b_{i\plus 1}}}
\put(0,-.2){\pp{t}{a_{i\mi 1}}}
\put(1,-.2){\pp{t}{a_{i}}}
\multiput(1.1,0)(.1,0){14}{\pp{}{.}}
\put(2,-.2){\pp{t}{a_{i}}}
\put(3,-.2){\pp{t}{a_{i\plus 1}}}
\end{picture}
   \label{eq:RowLambda}
\end{equation}

We decompose further the projector $P_{j+1}^{q+1}$ and sum over
$d_{i}$ to get the shorter fusion projector $P^{q}$:
\begin{equation}
     \setlength{\unitlength}{15pt}
     \frac{1}{S_{q}}\,
     \sum_{d_{i}\sim b_{i}}
\frac{\psi_{d_{i}}}{\psi_{b_{i}}}\;\;
\begin{picture}(6,2)(-1.5,1)
\multiput(0,0)(0,2){2}
{
\multiput(0,0)(3,-1){2}{\line(1,1){1}}
\multiput(0,0)(3,1){2}{\line(1,-1){1}}
\multiput(1,-1)(0,2){2}{\line(1,0){2}}
\put(2,0){\pp{}{P_{j\plus 2}^{q}}}
}
\put(-1,1){\line(1,1){1}}
\put(-1,1){\line(1,-1){1}}
\put(0,1){\pp{}{\sss-{(q-1)}\lambda}}
\put(0,.2){\pp{b}{\sss \to}}
\put(0,2){\pp{br}{b_{i}\;}}
\put(-1,1){\pp{r}{d_{i}\;}}
\put(0,0){\pp{tr}{b_{i}\;}}
\put(4,0){\pp{l}{\;a_{i}}}
\put(4,2){\pp{l}{\;a_{i}}}
\end{picture}
\;=\;
     \frac{S_{q\plus 1}}{S_{q}}\,
     \begin{picture}(5.5,2)(-1,0)
\multiput(0,0)(3,-1){2}{\line(1,1){1}}
\multiput(0,0)(3,1){2}{\line(1,-1){1}}
\multiput(1,-1)(0,2){2}{\line(1,0){2}}
\put(2,0){\pp{}{P_{j\plus 2}^{q}}}
\put(0,0){\pp{r}{b_{i}\;}}
\put(4,0){\pp{l}{\;a_{i}}}
\end{picture}
\vtop to 3\unitlength { } \; .
     \label{eq:Sumd}
\end{equation}

\noindent
so that \eqref{YqX3} reduces to the product of the scalar contribution
for each column times the matrix valued function
$\vec T_0^{q-1}$.

We now give the details of the contribution of each
column for each boundary condition.

\subsection{Functional equation on the torus}
\label{sec:FunctEqTorus}
Each horizontally $p$-fused bulk column in $\vec T_0^q\: \vec T_q^1$
brings a scalar factor of $s_q^p(u)$ when collapsed by the contractor.
Hence the $N$ bulk faces contribute to the $\vec T_{0}^{q-1}$ term as
\begin{equation}
f_q^p(u)= [s_q^p(u)]^N\,.
\end{equation}
The contribution of this same column to the $\vec T_{0}^{q+1}$ term
comes from the removal of the common scalar factors which appear in the
process of vertical fusion of the top $(p,1)$-fused face with the
larger $(p,q)$-fused face, yielding a $(p,q+1)$-fused face. The result
is $f^{p}_{q-1}(u)$.

Likewise, an $r$-type seam contributes in the same proportion but with
a shift in the spectral parameter and an adjustment in the common
scalar factors, yielding~\eqref{eq:SeamPhi}. It is easily checked that
the braid limit of such a factor simply vanishes, hence the $s$-type
seams don't contribute to the TBA equation and the same holds for
$\zeta$-type seams.

\subsection{Functional equations on the cylinder}
\label{sec:FunctEqCylinder}
As we discussed in Section~\ref{sec:BoundaryWeigths},
an $(r,a)$ boundary is the
combination of an $r$-seam and a $(1,a)$ boundary so we restrict
ourselves to $(1,a)$-boundary conditions.

In the cylinder case, the product $\vec T_0^q\: \vec T_q^1$ of double
row transfer matrices is realized as four layers of rows and a typical
column is the stack of two $(1,1)$-faces on top of two $(1,q)$-faces,
respectively at spectral parameters $\mu-u+\xi-q\lambda$,
$u+\xi+q\lambda$, $\mu-u+\xi-(q-1)\lambda$ and $u+\xi$ where
$\xi=-k\lambda$ for a usual bulk term involved in a horizontally fused
face.  Consider the following inversion
relation~\eqref{eq:InversionRelation}
\begin{gather}
     X_{j}^{1q}(2u-\mu+(2q-1)\lambda)
     X_{j}^{q1}(-2u+\mu-(2q-1)\lambda)=\qquad    \label{eq:Invers2u}
\\
   \qquad \qquad \qquad 
\;s_{2q}^{1\,q}(2u-\mu)\,s_{2-2q}^{1\,q}(-2u+\mu)\:P_{j}^{q+1}\,.
     \notag
\end{gather}
It follows from the
GYBE~\eqref{eq:GYBE} that
\begin{equation}
  \setlength{\unitlength}{10mm}
  \vtop to 3\unitlength{}
\topped{s_{2q}^{1\,q}{\sss(2u-\mu)}\times\;\;\;}{s_{2-2q}^{1\,q}{\sss 
(-2u+\mu)}}
\begin{picture}(1,3)(0,3)
     \put(0,0){\line(0,1){6}}
     \put(1,0){\line(0,1){6}}
     \put(0,0){\line(1,0){1}}
     \put(0,2){\line(1,0){1}}
\multiput(0,4)(0,1){3}	{
\put(0,0){\line(1,0){1}}
			}
\put(.5,1){\pp{}{u+\xi}}
\put(.5,3.2){\pp{}{\mu\mi u\plus\xi}}
\put(.5,2.8){\pp{}{\sss\mi (q-1)\lambda}}
\put(.5,4.7){\pp{}{u+\xi+}}
\put(.5,4.3){\pp{}{{q}\lambda}}
\put(.5,5.7){\pp{}{\mu\mi u}}
\put(.5,5.3){\pp{}{\plus\xi\mi{q}\lambda}}
\end{picture}
\;=\;
\begin{picture}(4.5,3)(-3.5,3)
     \put(0,0){\line(0,1){6}}
     \put(1,0){\line(0,1){6}}
     \put(0,0){\line(1,0){1}}
     \put(0,2){\line(1,0){1}}
\multiput(0,4)(0,1){3}	{
\put(0,0){\line(1,0){1}}
			}
\put(.5,1){\pp{}{u+\xi}}
\put(.5,3.2){\pp{}{\mu\mi u\plus\xi}}
\put(.5,2.8){\pp{}{\sss\mi (q-1)\lambda}}
\put(.5,4.7){\pp{}{u+\xi+}}
\put(.5,4.3){\pp{}{{q}\lambda}}
\put(.5,5.7){\pp{}{\mu\mi u}}
\put(.5,5.3){\pp{}{\plus\xi\mi{q}\lambda}}
\put(-2,2)	{
\put(0,1){\line(1,-1){1}}
\put(0,1){\line(1,2){1}}
\put(2,2){\line(-1,1){1}}
\put(2,2){\line(-1,-2){1}}
\put(1.2,1.7){\pp{}{-2u+\mu}}
\put(.8,1.3){\pp{}{-{\sss (2q-1)}\lambda}}
\put(.2,1.1){\pp{}{\downarrow}}
		}
\put(-4,2)	{
\put(0,2){\line(1,1){1}}
\put(0,2){\line(1,-2){1}}
\put(2,1){\line(-1,-1){1}}
\put(2,1){\line(-1,2){1}}
\put(.8,1.7){\pp{}{2u-\mu}}
\put(1.2,1.3){\pp{}{+{\sss (2q-1)}\lambda}}
\put(.2,2){\pp{}{\downarrow}}
		}
\multiput(-2.8,2)(.2,0){14}{\pp{}{.}}
\multiput(-2.8,5)(.2,0){14}{\pp{}{.}}
\put(-2,3){\pp{}{\bullet}}
\put(0,4){\pp{}{\bullet}}
\end{picture}
\;=\;
\begin{picture}(4.5,3)(-1.5,3)
     \put(0,0){\line(0,1){6}}
     \put(1,0){\line(0,1){6}}
     \put(0,0){\line(1,0){1}}
     \put(0,2){\line(1,0){1}}
     \put(0,3){\line(1,0){1}}
     \put(0,5){\line(1,0){1}}
     \put(0,6){\line(1,0){1}}
\put(.5,1){\pp{}{u+\xi}}
\put(.5,2.7){\pp{}{u+\xi+}}
\put(.5,2.3){\pp{}{{q}\lambda}}
\put(.5,4.2){\pp{}{\mu\mi u\plus\xi}}
\put(.5,3.8){\pp{}{\sss\mi (q-1)\lambda}}
\put(.5,5.7){\pp{}{\mu\mi u}}
\put(.5,5.3){\pp{}{\plus\xi\mi{q}\lambda}}
\put(1,2)	{
\put(0,1){\line(1,-1){1}}
\put(0,1){\line(1,2){1}}
\put(2,2){\line(-1,1){1}}
\put(2,2){\line(-1,-2){1}}
\put(1.2,1.7){\pp{}{-2u+\mu}}
\put(.8,1.3){\pp{}{-{\sss (2q-1)}\lambda}}
\put(.2,1.1){\pp{}{\downarrow}}
		}
\put(-2,2)	{
\put(0,2){\line(1,1){1}}
\put(0,2){\line(1,-2){1}}
\put(2,1){\line(-1,-1){1}}
\put(2,1){\line(-1,2){1}}
\put(.8,1.7){\pp{}{2u-\mu}}
\put(1.2,1.3){\pp{}{+{\sss (2q-1)}\lambda}}
\put(.2,2){\pp{}{\downarrow}}
		}
\multiput(-.8,2)(.2,0){4}{\pp{}{.}}
\multiput(-.8,5)(.2,0){4}{\pp{}{.}}
\multiput(1.2,2)(.2,0){4}{\pp{}{.}}
\multiput(1.2,5)(.2,0){4}{\pp{}{.}}
\put(0,3){\pp{}{\bullet}}
\put(1,3){\pp{}{\bullet}}
\end{picture}
\label{eq:CylinderColumn}
\end{equation}
As the rows on the right are of the same structure, the face
$X^{q1}(-2u+\mu-(2q-1)\lambda)$ can push through rightward all the way
to the right boundary.  Similarly, its counterpart
$X^{1q}(2u-\mu+(2q-1)\lambda)$ can push through leftwards all the way
to the left boundary.  Because of equation~\eqref{eq:Spectators},
these rectangular weights, after a crossing
symmetry~\eqref{eq:CrossingSymmetry}, simply agglomerate into a larger
boundary (minus the larger projector). For a $(1,a_{R})$-boundary
condition on the right hand side, it reads:

\begin{equation}
\setlength{\unitlength}{6mm}
\vtop to 5.2\unitlength{}
\begin{picture}(3.5,4.5)
%
%
\multiput(2,0)(2,0){2}{\pp{}{\bullet}}
\multiput(3,1)(2,0){1}{\pp{}{\bullet}}
\multiput(3,-1)(2,0){1}{\pp{}{\bullet}}
\multiput(4,2)(2,0){1}{\pp{}{\bullet}}
\multiput(4,-2)(2,0){1}{\pp{}{\bullet}}
\put(0,0){\line(1,1){5}}
\put(0,0){\line(1,-1){5}}
\put(1,1){\line(1,-1){4}}
\put(2,2){\line(1,-1){3}}
\put(3,3){\line(1,-1){2}}
\put(4,4){\line(1,-1){1}}
\put(1,-1){\line(1,1){4}}
\put(2,-2){\line(1,1){3}}
\put(3,-3){\line(1,1){2}}
\put(4,-4){\line(1,1){1}}
%
%
\put(4.05,3.35){\pp{c}{2u+}}\put(4,2.8){\pp{c}{2q\lambda-\mu}}
\put(3,2.4){\pp{c}{2u+}}\put(3,1.95){\pp{c}{(2q-1)\!
\lambda}}\put(3,1.6){\pp{c}{-\mu}}
\put(4.05,1.45){\pp{c}{2u+}}\put(4.05,1){\pp{c}{(2q-2)\!
\lambda}}\put(4.05,0.6){\pp{c}{-\mu}}
\put(1,0.45){\pp{c}{2u+}}\put(1,-0.05){\pp{c}{(q+1)
\lambda}}\put(1,-0.5){\pp{c}{-\mu}}
\put(4,-2.75){\pp{c}{2u+}}\put(4,-3.2){\pp{c}{2
\lambda-\mu}}
\multiput(0.75,-0.9)(1,-1){4}{\pp{bl}{\rightarrow}}
\multiput(1.75,0.1)(1,-1){3}{\pp{bl}{\rightarrow}}
\multiput(2.75,1.1)(1,-1){2}{\pp{bl}{\rightarrow}}
\multiput(3.75,2.1)(1,-1){1}{\pp{bl}{\rightarrow}}
\multiput(5.7,-3.2)(0,2){5}{\pp{r}{a_R}}
\multiput(5.7,-5.2)(0,2){1}{\pp{r}{a_R}}
\multiput(5,-5)(0,0.2){50}{\pp{}{.}}
\put(0,0){\pp{r}{c\;\;}}
\put(-4,0){\p{}{\ds \frac{\psi_{c}^{\half}}{\psi_{a_{r}}^{\half}}
\prod_{i=1}^{q}\frac{1}{s^{i1}_{-i}(\mu-2u)}}}
\end{picture}
\label{eq:Aglommerat}
\end{equation}

In the bulk, we are now in a similar configuration as in the torus
case, simply the columns are doubled.  Hence the same technique
applies provided the push through of the contractor $X(\lambda)$ and
the longer projector $P^{q+2}$ behave as expected on the boundary.  This
is what we discuss next.

It is clear from~\eqref{eq:Aglommerat} that the push-through property
of projectors is still satisfied on the boundary, the projector
$P^{q+1}$ coming from the bottom $q$ rows of the product can go up
through the right boundary and get back to the $q$ intermediate rows,
lower part of the top half.  Similarly, coming from the right of these
rows, it can go down the left boundary to the lower $q$ rows.
Therefore, the term proportional $\vec T^{q+1}_{0}$ proceeds in
exactly the same way as in the case of the torus.

%
%
Consider now the term $\vec T^{q-1}_{0}$.  We need to understand the
action of $X_j(\lambda)$ on the $(1,a_R)$ right boundary.

Similarly to \eqref{collapse1}, we have
\begin{equation}
\setlength{\unitlength}{6mm}
\begin{picture}(8,3)
\put(-4,0.5){
\begin{picture}(3.5,3.5)
\put(0,1){\line(1,1){1}}
\put(0,1){\line(1,-1){1}}
\put(2,1){\line(-1,1){1}}
\put(2,1){\line(-1,-1){1}}
\multiput(2,0)(1.3,0){2}{\line(0,1){2}}
\multiput(2,0)(0,1){3}{\line(1,0){1.3}}
\put(0.2,1){\pp{}{\downarrow}}
\multiput(2.2,0.26)(0,1){2}{\pp{}{\searrow}}
\put(1,1){\pp{}{\lambda}}
\put(2.7,0.5){\pp{c}{u+\lambda}}
\put(2.7,1.5){\pp{c}{u}}
\multiput(1,0)(0.1,0){10}{\pp{}{.}}
\multiput(1,2)(0.1,0){10}{\pp{}{.}}
\put(2,1){\pp{}{\bullet}}
\put(-0.27,1.1){\pp{c}{d_{i}}}
\put(1,-0.2){\pp{c}{a_{i}}}
\put(2,-0.2){\pp{c}{a_{i}}}
\put(1,2.2){\pp{c}{b_{i}}}
\put(2,2.2){\pp{c}{b_{i}}}
\put(3.4,-0.2){\pp{c}{a_{i\!+\!1}}}
\put(3.4,2.2){\pp{c}{b_{i\!+\!1}}}
\put(3.75,1){\pp{c}{d_{i\!+\!1}}}
\end{picture} }
\put(2,1){$\displaystyle=$}
\put(3.2,1){$\displaystyle
s_1(u)s_1(-u)
$}
\put(8,0.5){
\begin{picture}(3.5,3.5)
\put(-0.3,0){\line(1,0){1.3}}
\put(-0.3,0){\line(0,1){1}}
\put(1,1){\line(-1,0){1.3}}
\put(1,1){\line(0,-1){1}}
\put(1,1){\line(1,1){1}}
\put(1,1){\line(1,-1){1}}
\put(3,1){\line(-1,1){1}}
\put(3,1){\line(-1,-1){1}}
\multiput(1,0)(0.1,0){10}{\pp{}{.}}
\put(-0.3,-0.2){\pp{c}{a_{i}}}
\put(1,-0.2){\pp{c}{a_{i\!+\!1}}}
\put(2.2,-0.2){\pp{c}{a_{i\!+\!1}}}
\put(-0.3,1.25){\pp{c}{d_{i}}}
\put(0.9,1.25){\pp{c}{b_{i}}}
\put(2,2.25){\pp{c}{b_{i\!+\!1}}}
\put(3.45,1){\pp{c}{d_{i\!+\!1}}}
\put(0,0.8){\pp{}{\swarrow}}
\put(1.35,1){\pp{}{\downarrow}}
\put(0.35,0.5){\pp{c}{\lambda}}
\put(2,1){\pp{c}{\lambda}}
\end{picture} }
\end{picture}
\label{collapse2}
\end{equation}
So that when the contractor acts on the $(1,a_R)$ right boundary, we
get
\begin{equation}
\setlength{\unitlength}{6mm}
\begin{picture}(8,10.5)
\put(-3,5.5){
\begin{picture}(3.5,5.5)
%
\put(0,0){\thicklines\line(-1,0){1.5}}
\multiput(0,-5)(0,1){10}{\line(-1,0){1.5}}
\put(-1.2,-5){\line(0,1){9}}
\put(0,-5){\line(0,1){9}}
\multiput(0,1)(0.2,0){5}{\pp{}{.}}
\multiput(0,-1)(0.2,0){5}{\pp{}{.}}
\multiput(0,2)(0.2,0){10}{\pp{}{.}}
\multiput(0,-2)(0.2,0){10}{\pp{}{.}}
\multiput(0,3)(0.2,0){15}{\pp{}{.}}
\multiput(0,-3)(0.2,0){15}{\pp{}{.}}
\multiput(0,4)(0.2,0){10}{\pp{}{.}}
\multiput(0,-4)(0.2,0){20}{\pp{}{.}}
\multiput(0,-5)(0.2,0){25}{\pp{}{.}}
\multiput(-1,-4.8)(0,1){8}{\pp{}{
\searrow }}
\put(-0.85,3.8){\pp{}{\swarrow}}
\put(0.2,4.2){\pp{c}{b_{N\!-\!1} }}
\put(-1.2,4.2){\pp{c}{b_{N\!-\!2} }}
\put(-1.75,3.2){\pp{c}{a_{N\!-\!1} }}
\put(0.45,3.2){\pp{c}{a_{N} }}
\put(-0.5,-4.5){\pp{c}{u}}
\put(-0.5,-0.5){\pp{c}{u\!+\!q\lambda}}
\put(-0.6,-1.2){\pp{c}{u\!+}}
\put(-0.5,-1.6){\pp{c}{(\!q\!-\!1\!)\!\lambda}}
\put(-0.5,3.5){\pp{c}{\lambda}}
\put(-0.6,0.5){\pp{c}{-u\!+\!\mu}}
\put(-0.55,2.75){\pp{c}{\!-u\!+\!\mu}}
\put(-0.4,2.3){\pp{c}{-\!(\!q\!-\!2\!)\!\lambda}}
\multiput(0,0)(0,1){8}{
\multiput(0,-4)(-1.2,0){2}{\pp{}{\bullet}} }
%
\multiput(0,0)(2,0){3}{\pp{}{\bullet}}
\multiput(1,1)(2,0){2}{\pp{}{\bullet}}
\multiput(1,-1)(2,0){2}{\pp{}{\bullet}}
\multiput(2,2)(2,0){2}{\pp{}{\bullet}}
\multiput(2,-2)(2,0){2}{\pp{}{\bullet}}
\multiput(3,3)(2,0){1}{\pp{}{\bullet}}
\multiput(3,-3)(2,0){1}{\pp{}{\bullet}}
\multiput(4,4)(2,0){1}{\pp{}{\bullet}}
\multiput(4,-4)(2,0){1}{\pp{}{\bullet}}
\put(0,0){\line(1,1){5}}
\put(0,0){\line(1,-1){5}}
\put(1,1){\line(1,-1){4}}
\put(2,2){\line(1,-1){3}}
\put(3,3){\line(1,-1){2}}
\put(4,4){\line(1,-1){1}}
\put(1,-1){\line(1,1){4}}
\put(2,-2){\line(1,1){3}}
\put(3,-3){\line(1,1){2}}
\put(4,-4){\line(1,1){1}}
%
%
\put(4.05,3.35){\pp{c}{2u+}}\put(4,2.8){\pp{c}{2q\lambda-\mu}}
\put(3,2.4){\pp{c}{2u+}}\put(3,1.95){\pp{c}{(2q-1)\!
\lambda}}\put(3,1.6){\pp{c}{-\mu}}
\put(4.05,1.45){\pp{c}{2u+}}\put(4.05,1){\pp{c}{(2q-2)\!
\lambda}}\put(4.05,0.6){\pp{c}{-\mu}}
\put(1,0.45){\pp{c}{2u+}}\put(1,-0.05){\pp{c}{(q+1)
\lambda}}\put(1,-0.5){\pp{c}{-\mu}}
\put(4,-2.75){\pp{c}{2u+}}\put(4,-3.2){\pp{c}{2
\lambda-\mu}}
\put(3,3){\line(-1,1){1}}
\put(2,4){\line(1,1){1}}
\put(4,4){\line(-1,1){1}}
\put(3,4){\pp{c}{\lambda}}
\multiput(0.75,-0.9)(1,-1){4}{\pp{bl}{\rightarrow}}
\multiput(1.75,0.1)(1,-1){3}{\pp{bl}{\rightarrow}}
\multiput(2.75,1.1)(1,-1){2}{\pp{bl}{\rightarrow}}
\multiput(3.75,2.1)(1,-1){1}{\pp{bl}{\rightarrow}}
\multiput(2.25,4.03)(-1,1){1}{\pp{}{\downarrow}}
\multiput(5.7,-3.2)(0,2){5}{\pp{r}{a_R}}
\multiput(5.7,-5.2)(0,2){1}{\pp{r}{a_R}}
\multiput(2.55,5.1)(0,-1.35){1}{\pp{l}{a_R}}
\multiput(5,-5)(0,0.2){50}{\pp{}{.}}
\multiput(5,5)(-0.2,0){10}{\pp{}{.}}
\end{picture} }
\put(3,5.5){$\displaystyle =$}
\put(6.5,5.5){
\begin{picture}(3.5,5.5)
%
\put(0,0){\thicklines\line(-1,0){1.5}}
\multiput(0,-5)(0,1){10}{\line(-1,0){1.5}}
\put(-1.2,-5){\line(0,1){9}}
\put(0,-5){\line(0,1){9}}
\multiput(0,1)(0.2,0){5}{\pp{}{.}}
\multiput(0,-1)(0.2,0){5}{\pp{}{.}}
\multiput(0,2)(0.2,0){10}{\pp{}{.}}
\multiput(0,-2)(0.2,0){10}{\pp{}{.}}
\multiput(0,3)(0.2,0){15}{\pp{}{.}}
\multiput(0,-3)(0.2,0){15}{\pp{}{.}}
\multiput(0,4)(0.2,0){10}{\pp{}{.}}
\multiput(0,-4)(0.2,0){20}{\pp{}{.}}
\multiput(0,-5)(0.2,0){25}{\pp{}{.}}
\multiput(-1,-4.8)(0,1){8}{\pp{}{
\searrow }}
\put(-0.85,3.8){\pp{}{\swarrow}}
\put(0.2,4.2){\pp{c}{b_{N\!-\!1} }}
\put(-1.2,4.2){\pp{c}{b_{N\!-\!2} }}
\put(-1.75,3.2){\pp{c}{a_{N\!-\!1} }}
\put(0.45,3.2){\pp{c}{a_{N} }}
\put(-0.5,-4.5){\pp{c}{u}}
\put(-0.5,-0.5){\pp{c}{u\!+\!q\lambda}}
\put(-0.6,-1.2){\pp{c}{u\!+}}
\put(-0.5,-1.6){\pp{c}{(\!q\!-\!1\!)\!\lambda}}
\put(-0.5,3.5){\pp{c}{\lambda}}
\put(-0.6,0.5){\pp{c}{-u\!+\!\mu}}
\put(-0.55,2.75){\pp{c}{\!-u\!+\!\mu}}
\put(-0.4,2.3){\pp{c}{-\!(\!q\!-\!2\!)\!\lambda}}
\multiput(0,0)(0,1){8}{
\multiput(0,-4)(-1.2,0){2}{\pp{}{\bullet}} }
%
\multiput(0,0)(2,0){3}{\pp{}{\bullet}}
\multiput(1,1)(2,0){2}{\pp{}{\bullet}}
\multiput(1,-1)(2,0){2}{\pp{}{\bullet}}
\multiput(2,2)(2,0){2}{\pp{}{\bullet}}
\multiput(2,-2)(2,0){2}{\pp{}{\bullet}}
\multiput(3,3)(2,0){1}{\pp{}{\bullet}}
\multiput(3,-3)(2,0){1}{\pp{}{\bullet}}
\multiput(4,4)(2,0){1}{\pp{}{\bullet}}
\multiput(4,-4)(2,0){1}{\pp{}{\bullet}}
\put(0,0){\line(1,1){5}}
\put(0,0){\line(1,-1){5}}
\put(1,1){\line(1,-1){4}}
\put(2,2){\line(1,-1){3}}
\put(3,3){\line(1,-1){2}}
\put(4,4){\line(1,-1){1}}
\put(1,-1){\line(1,1){4}}
\put(2,-2){\line(1,1){3}}
\put(3,-3){\line(1,1){2}}
\put(4,-4){\line(1,1){1}}
%
%
\put(4,3){\pp{c}{\lambda}}
\put(3,2.4){\pp{c}{2u+}}\put(3,1.95){\pp{c}{(2q-1)\!
\lambda}}\put(3,1.6){\pp{c}{-\mu}}
\put(4.05,1.45){\pp{c}{2u+}}\put(4.05,1){\pp{c}{(2q-2)\!
\lambda}}\put(4.05,0.6){\pp{c}{-\mu}}
\put(1,0.45){\pp{c}{2u+}}\put(1,-0.05){\pp{c}{(q+1)
\lambda}}\put(1,-0.5){\pp{c}{-\mu}}
\put(4,-2.75){\pp{c}{2u+}}\put(4,-3.2){\pp{c}{2
\lambda-\mu}}
\put(3,3){\line(-1,1){1}}
\put(2,4){\line(1,1){1}}
\put(4,4){\line(-1,1){1}}
\put(3.1,4.35){\pp{c}{2u+}}
\put(3,3.85){\pp{c}{2q\lambda-\mu}}
\multiput(0.75,-0.9)(1,-1){4}{\pp{bl}{\rightarrow}}
\multiput(1.75,0.1)(1,-1){3}{\pp{bl}{\rightarrow}}
\multiput(2.75,1.1)(1,-1){2}{\pp{bl}{\rightarrow}}
\multiput(3.75,2.1)(1,-1){1}{\pp{bl}{\rightarrow}}
\multiput(2.25,4.03)(-1,1){1}{\pp{}{\downarrow}}
\multiput(5.7,-3.2)(0,2){5}{\pp{r}{a_R}}
\multiput(5.7,-4.75)(0,2){1}{\pp{r}{a_R}}
\multiput(2.55,5.1)(0,-1.35){1}{\pp{l}{a_R}}
\multiput(5,-5)(0,0.2){50}{\pp{}{.}}
\multiput(5,5)(-0.2,0){10}{\pp{}{.}}
\end{picture} }
\end{picture}
\label{VBinitial}
\end{equation}
where we have used the crossing symmetry and
then the abelian property \eqref{eq:Abelian} to interchange
the parameters $\lambda$ and $2u+2q\lambda-\mu$ between
two face weights. Then, we apply \eqref{collapse1} to
collapse the faces weights inside the $(1,a_R)$ boundary

\begin{equation}
\setlength{\unitlength}{6mm}
\begin{picture}(8,10)
\put(-6.9,6){$\displaystyle =$}
\put(-5.5,6){$\displaystyle
\begin{array}{c}
  (-1)^{q-1} s_{2q+1}(2u-\mu)\times \\
  s_{2q-3}^{q-1}(2u-\mu)
  s_{2q-1}^{q-1}(2u-\mu)
\end{array}
$}
\put(5,6){
\begin{picture}(3.5,5.5)
%
\put(0,0){\thicklines\line(-1,0){1.5}}
\multiput(0,-5)(0,1){10}{\line(-1,0){1.5}}
\put(-1.2,-5){\line(0,1){9}}
\put(0,-5){\line(0,1){9}}
\multiput(0,1)(0.2,0){5}{\pp{}{.}}
\multiput(0,-1)(0.2,0){5}{\pp{}{.}}
\multiput(0,2)(0.2,0){10}{\pp{}{.}}
\multiput(0,-2)(0.2,0){10}{\pp{}{.}}
\multiput(0,3)(0.2,0){15}{\pp{}{.}}
\multiput(0,-3)(0.2,0){15}{\pp{}{.}}
\multiput(0,4)(0.2,0){20}{\pp{}{.}}
\multiput(0,-4)(0.2,0){20}{\pp{}{.}}
\multiput(0,-5)(0.2,0){25}{\pp{}{.}}
\multiput(-1,-4.8)(0,1){8}{\pp{}{
\searrow }}
\put(-0.85,3.8){\pp{}{\swarrow}}
\put(0.2,4.2){\pp{c}{b_{N\!-\!1} }}
\put(-1.2,4.2){\pp{c}{b_{N\!-\!2} }}
\put(-1.75,3.2){\pp{c}{a_{N\!-\!1} }}
\put(0.45,3.2){\pp{c}{a_{N} }}
\put(-0.5,-4.5){\pp{c}{u}}
\put(-0.5,-0.5){\pp{c}{u\!+\!q\lambda}}
\put(-0.6,-1.2){\pp{c}{u\!+}}
\put(-0.5,-1.6){\pp{c}{(\!q\!-\!1\!)\!\lambda}}
\put(-0.5,3.5){\pp{c}{\lambda}}
\put(-0.6,0.5){\pp{c}{-u\!+\!\mu}}
\put(-0.55,2.75){\pp{c}{\!-u\!+\!\mu}}
\put(-0.4,2.3){\pp{c}{-\!(\!q\!-\!2\!)\!\lambda}}
\multiput(0,0)(0,1){8}{
\multiput(0,-4)(-1.2,0){2}{\pp{}{\bullet}} }
%
\multiput(0,0)(2,0){3}{\pp{}{\bullet}}
\multiput(1,1)(2,0){2}{\pp{}{\bullet}}
\multiput(1,-1)(2,0){2}{\pp{}{\bullet}}
\multiput(2,2)(2,0){2}{\pp{}{\bullet}}
\multiput(2,-2)(2,0){2}{\pp{}{\bullet}}
\multiput(3,3)(2,0){1}{\pp{}{\bullet}}
\multiput(3,-3)(2,0){1}{\pp{}{\bullet}}
\multiput(4,-4)(2,0){1}{\pp{}{\bullet}}
\put(4.1,4.2){\pp{c}{b_{N\!-\!1} }}
\put(2.9,3.3){\pp{c}{a_{N} }}
\put(0,0){\line(1,1){4}}
\put(0,0){\line(1,-1){5}}
\put(1,1){\line(1,-1){1}}
\put(2,2){\line(1,-1){1}}
\put(3,-1){\line(1,-1){2}}
\put(4,0){\line(1,-1){1}}
\put(3,3){\line(1,-1){1}}
\put(4,4){\line(1,-1){1}}
\put(1,-1){\line(1,1){4}}
\put(2,-2){\line(1,1){3}}
\put(3,-3){\line(1,1){2}}
\put(4,-4){\line(1,1){1}}
%
%
%
\put(4,3){\pp{c}{\lambda}}
\put(3,2){\pp{c}{\lambda}}
\put(2,1){\pp{c}{\lambda}}
\put(1,0){\pp{c}{\lambda}}
\put(4,-2.75){\pp{c}{2u+}}\put(4,-3.2){\pp{c}{2
\lambda-\mu}}
\multiput(0.75,-0.9)(1,-1){1}{\pp{bl}{\rightarrow}}
\multiput(2.75,-2.9)(1,-1){2}{\pp{bl}{\rightarrow}}
\multiput(1.75,0.1)(1,-1){1}{\pp{bl}{\rightarrow}}
\multiput(3.75,-1.9)(1,-1){1}{\pp{bl}{\rightarrow}}
\multiput(2.75,1.1)(1,-1){1}{\pp{bl}{\rightarrow}}
\multiput(3.75,2.1)(1,-1){1}{\pp{bl}{\rightarrow}}
\multiput(5.7,-3.2)(0,2){4}{\pp{r}{a_R}}
\multiput(5.7,-5.2)(0,2){1}{\pp{r}{a_R}}
\multiput(5,-5)(0,0.2){40}{\pp{}{.}}
\multiput(2,0)(0,-0.2){10}{\pp{}{.}}
\multiput(3,1)(0,-0.2){10}{\pp{}{.}}
\multiput(4,2)(0,-0.2){10}{\pp{}{.}}
\end{picture} }
\end{picture}
\label{VacuumDown}
\end{equation}
where 
we use the identity~\eqref{eq:XP}
to eliminate the face weight with
parameter $2u+2q\lambda-\mu$.

We can see in \eqref{VacuumDown}, that the right $(1,a_R)$ boundary of
$\vec T_0^q\:\vec T_1^1$ is contracted into a smaller $B_R^{q-1}$
under the action of the contractor $X_j(\lambda)$.  We then continue
to collapse the face weights in the bottom half of the transfer matrix
with the contractor.

For the $(1,a_L)$ left boundary, the contractor $X_j(\lambda)$ acts
from the bottom.  We rotate the whole diagram by half a turn and we
apply the same technique, and we get the following scalar factors
\begin{equation}
(-1)^q s_{2q-3}(2u-\mu) s_{2q-2}^{q-1}(2u-\mu)
s_{2q-4}^{q-1}(2u-\mu)\, .
\end{equation}

Finally, the contractor can go back to the top row from the left,
hence the rest of the proof proceeds as previously.

Collecting the different contributions for all the columns, which come
in pairs, for the lower and the upper halves of the product $\vec
T_0^q\:\vec T_1^1$, with spectral parameters $u+\xi$ and
$\mu-u-q\lambda+\xi$ respectively, one gets the result listed in
Section~\ref{sec:FusionHierarchies}.

\section{Discussion}
\label{sec:Discussion}\setcounter{equation}{0}
In this paper we have derived the TBA functional equations for
critical lattice models using simple fusion projectors.  We point out,
however, that the very same functional equations can be derived
off-criticality by using the methods of \cite{BPO'B96}.  This applies,
for example, for the $A$ and $D$ models which admit elliptic solutions
to the Yang-Baxter equations.

We conjecture generally that the form of the TBA functional equations
are universal for all integrable lattice models associated with
rational CFTs and their integrable perturbations.  In particular, we
expect the known forms~\cite{KNS1} 
of these equations to
apply to all integrable boundary conditions.

\section*{Acknowledgements}
\label{sec:Acknowldegments}
This research is supported by the Australian Research Council.
%
\nobreak
\let\chapter\section 
\bibliographystyle{unsrtnat} 
\bibliography{seam}
\end{document}